\documentclass[aps,preprint,epsfig,rotate]{revtex4}
\usepackage{tikz}
\usepackage{verbatim}
\usepackage{sansmath,pdflscape,graphicx,rotating}
\usetikzlibrary{calc,arrows,decorations.pathmorphing,intersections}
\usepackage[font={small,sf},labelfont={bf},labelsep=endash]{caption}

\begin{document}
%\begin{doublespace}

\title{Singlet and triplet bound state spectra in the four-electron Be-like atomic systems}

\author{Mar\'{\i}a Bel\'en Ruiz}
\email[E--mail address: ]{maria.belen.ruiz@fau.de}

\affiliation{Department of Theoretical Chemistry, \\
Friedrich-Alexander-University Erlangen-N\"urnberg, Egerlandstra\ss e 3,
D-91058, Erlangen, Germany}

\author{Federico Latorre}
\email[E--mail address: ] {latorrf3@univie.ac.at}

\affiliation{Department of Theoretical Chemistry, \\
Friedrich-Alexander-University Erlangen-N\"urnberg, Egerlandstra\ss e 3,
D-91058, Erlangen, Germany}

\affiliation{Institute of Theoretical Chemistry, University of Vienna, 
W\"ahringer Str. 17, 1090 Vienna, Austria}

\author{Alexei M. Frolov}
\email[E--mail address: ]{afrolov@uwo.ca}

\affiliation{Department of Applied Mathematics \\
 University of Western Ontario, London, Ontario N6H 5B7, Canada}

\date{\today}

\begin{abstract}
A large number of bound singlet and triplet states in the four-electron Be-atom and Be-like ions B$^+$, C$^{2+}$,
F$^{5+}$ and Mg$^{8+}$ are determined to milli-Hartree numerical accuracy ($1\cdot 10^{-3}$ a.u.).
These states include the bound singlet and triplet $S$-, $P$-, $D$-, $F$-, $G$-, $H$-, $I$- and $K$-states. Based on
computational results we analyze the singlet and triplet series of the Be atom bound state spectrum and some
four-electron ions: B$^+$, C$^{2+}$,
F$^{5+}$ and Mg$^{8+}$ (Be-like ions). The analogous
study of the Be atom triplet states was the topic of our earlier study
(A.M. Frolov and M.B. Ruiz, CPL
\textbf{595-596}, 197 (2014)). The computational data allowed us to draw the spectral diagram of the bound state
spectrum of the Be atom and other four-electron ions mentioned above. The Be atom spectrum contains the
two optical series of bound states: singlets and triplets.

\vspace{1cm}


\end{abstract}

\maketitle

\newpage

\section{Introduction}

The electronic structure of the beryllium atom is of great interest in various problems arising in different areas of
modern science, including stellar astrophysics and plasmas, high-temperature physics and applied nuclear
physics. Beryllium and some of its compounds (BeO, Be$_{2}$C) are extensively used in the nuclear industry mainly as
very effective (almost ideal) moderators of fast/slow neutrons. Nevertheless, currently there are many gaps in
our understanding of the Be-atom optical spectrum. Total energies of all rotationally excited
(bound) states with $L \ge 4$ in particular have not yet been evaluated, to our knowledge.

Another interesting problem is to describe the
transitions from the spectra of the low-lying bound states in the Be atom to the weakly-bound (or Rydberg) states.

Recently, we have studied the general structure of the triplet bound state spectrum in the four-electron Be-atom \cite{Our1} 
and accurately calculated a large number of low-lying (bound) $S$-, $P$-,
$D$-, $F$-, $G$-, $H$-, $I$- and $K$-states, i.e. bound states with $L \le 7$. These computational results
allowed us to determine a spectral diagram of the triplet states of the Be atom. The theoretical/computational
spectral diagram of the Be-atom agrees well with the known experimental data (see, e.g., \cite{Kramida} and references therein). 
N.B. high angular momentum states e.g. $G(L=4)$, $H(L=5)$, $I(L=6)$, and $K(L=7)$ are absent from atomic data bases \cite{Kramida}.

Since \cite{SH71} the bound states in the four-electron Be-atom and Be-like ions have been considered by many authors, using
various highly accurate methods specifically designed for the four-electron atomic systems. Such calculations were
restricted to the ground $2^1S$-state and a very few excited states only. For instance, the ground state energy
for the Be-atom was determined by applying the Configuration Interaction (CI) method with Slater type orbitals (STO) \cite{Bunge,Chung}, the
Hylleraas method (Hy) \cite{Luechow,King}, the Hylleraas-Configuration Interaction method (Hy-CI) \cite{SH71,SH2011}
and the Exponential Correlated Gaussian (ECG) method \cite{PKP13,AdamS} (also called variational expansion in multi-dimensional gaussoids \cite{KT}).
A few selected bound singlet $S$-, $P$- and $D$-states were calculated by Monte Carlo methods \cite{Galv}
and by the ECG method \cite{PKP13,AdamS,AdamP,AdamD}. In contrast to singlet states, the Be-atom triplet states were
investigated only in a very few earlier studies \cite{Bertini,Galv,Frolov2009} (see also \cite{Our1} and references therein).

To complete the analysis of the bound state spectrum of the Be-atom \cite{Our1} we need to consider the singlet series of bound
states in this system. In this study we determine the total energies of the bound singlet $S$-, $P$-, $D$-, $F$-, $G$-, $H$-, $I$-
and $K$-states. 

The method of calculation allows the total energies of these states to be obtained to very high numerical accuracy,
$\approx$ few milli-Hartree ($1\cdot 10^{-3}$ a.u.), which is
significantly better (for highly excited states) than accuracy which Hartree-Fock based methods can provide. It is important to note,
that currently there is no reliable experimental information about rotationally excited states with $L \ge 4$ in the Be-atom. Therefore,
the main goal of this study is to determine the bound state spectrum of low-lying singlet states in the four-electron Be-atom, including
rotationally excited states with $L$ = 4, 5, 6 and 7. The results of this study have been represented as a spectral
diagram for the singlet and triplet bound states in the Be-atom. For readers benefit and to complete our analysis we also present an
analogous spectral diagram for the bound triplet states in the Be atom \cite{Our1}.

\section{Hamiltonian and bound state wave functions in CI-method}

The computational goal of this study is to determine the accurate numerical solutions of the five-body (or four-electron)
Schr\"{o}dinger equation $H \Psi = E \Psi$, where the Hamiltonian written in Hylleraas coordinates for a CI wave function is written in
the form (see, e.g., \cite{Ruiz})
\begin{eqnarray}
 \hat{H} &=&-\frac 12\sum_{i=1}^n\frac{\partial ^2}{\partial r_i^2}
- \sum_{i=1}^n\frac 1{r_i}\frac \partial {\partial r_i}-\sum_{i=1}^n\frac
 Z{r_i}+\sum_{i<j}^n\frac 1{r_{ij}} \nonumber \\
 &-&\frac 12\sum_{i=1}^n\frac 1{r_i^2}\frac{\partial ^2}{\partial \theta _i^2}
 - \frac 12\sum_{i=1}^n\frac 1{r_i^2\sin ^2{\theta _i}}\frac{\partial ^2}{%
\partial \varphi _i^2}-\frac 12\sum_{i=1}^n\frac{\cot {\theta _i}}{r_i^2}%
\frac \partial {\partial \theta _i}  \label{eq1}
\end{eqnarray}
Note that when the Hamiltonian in Hylleraas coordinates is applied to the CI wave function (the CI wave function does not explicitly include
$r_{ij}$ coordinates) some terms of \cite{Ruiz} vanish. In addition, the kinetic energy operator is represented as a sum of a few terms
each of which has its own radial and angular parts. The operator $\hat{H}$ is diagonal in the basis of the spherical harmonics which are used
below as angular parts of the orbital functions. Note also that the angular momentum operator can easily be extracted from Eq.(\ref{eq1}):
\begin{equation}
\sum_{i=1}^n\frac 1{r_i^2}\hat{L}_i^2=-\frac 12\sum_{i=1}^n\frac 1{r_i^2}%
\frac{\partial ^2}{\partial \theta _i^2}-\frac 12\sum_{i=1}^n\frac
1{r_i^2\sin ^2{\theta _i}}\frac{\partial ^2}{\partial \varphi _i^2}-\frac
12\sum_{i=1}^n\frac{\cot {\theta _i}}{r_i^2}\frac \partial {\partial \theta
_i},
\end{equation}
For the orbital basis functions $\phi _i$ (or orbitals, for short) we can write
\begin{equation}
L_i^2\phi _i=l_i(l_i+1)\phi _i,
\end{equation}
with $l_i$ the angular quantum number of the orbital $\phi _i$. The Hamiltonian is reduced to the form

\begin{eqnarray}
\hat{H} &=&-\frac 12\sum_{i=1}^n\frac{\partial ^2}{\partial r_i^2}%
-\sum_{i=1}^n\frac 1{r_i}\frac {\partial}{\partial r_i}-\sum_{i=1}^n\frac
{Z}{r_i}+\sum_{i<j}^n\frac 1{r_{ij}} + \sum_{i=1}^n\frac 1{r_i^2} l_i(l_i+1)
\end{eqnarray}

Now, from the variational principle one obtains the following eigenvalue problem:

\begin{equation}
(\mathbf{H}-E\mathbf{S})\mathbf{C}=\mathbf{0,}
\end{equation}
where the matrix elements of the Hamiltonian matrix $\mathbf{H}$ and overlap matrix $\mathbf{S}$ are:

\begin{equation}
H_{kl}=\int \Phi _kH\Phi _ld\tau ,\text{ \qquad }S_{kl}=\int \Phi _k\Phi
_ld\tau .
\end{equation}

The integrals occurring in the CI calculations of an $n$-electron atom are one- and two-electron integrals. The two-electron integrals are
of the type \cite{Ruiz2e}
\begin{equation}
\left\langle
\phi (\mathbf{r}_1) \phi (\mathbf{r}_2) \frac 1{r_{12}} \phi (\mathbf{r}_1) \phi (%
\mathbf{r}_2) \right\rangle
\end{equation}
and they are expressed as sums of the auxiliary two-electron integrals $V(m,n;\alpha ,\beta )$, defined as:

\begin{equation}
  V(m,n;\alpha ,\beta )=\int_0^\infty r_1^me^{-\alpha r_1}dr_1\int_{r_1}^\infty r_2^ne^{-\beta r_2}dr_2\ ,
\end{equation}

The auxiliary integrals $V(m,n;\alpha ,\beta )$ for positive indices $m,n$
consist on a sum of $A(n,\alpha )$ auxiliary integrals \cite[Eq. (5)]{Frolov-A}:
\begin{equation}
V(m,n;\alpha ,\beta )=\sum_{n^{\prime }=0}^n{n \choose {n^{\prime }}}%
A(n^{\prime },\alpha )A(m+n-n^{\prime },\alpha +\beta ),\qquad m,n\geq 0
\end{equation}
This formula, developed by Frolov and Smith, is very useful because it is numerically stable and provides very fast convergence.
For negative $n$ and positive $m$ (but $m+n\geq -1$ always) the formula for the Sims and Hagstrom sum \cite[Eq. (33)]{Sims04} must be computed. In
quantum mechanical calculations of two-electron systems this formula is employed to calculate the V-auxiliary
integrals with the lowest index $n=-1$:
\begin{equation}
V(m,n;\alpha ,\beta )=\sum_{q=1}^\infty \frac{\alpha ^{q-1}m!}{(m+q)!}%
A(m+n+q;\alpha +\beta ),\qquad m+n\geq -1,\text{ \ }m>0,\text{ \ \ }n<0
\end{equation}
where the $A(n,\alpha )$ auxiliary integrals are:
\begin{equation}
A(n,\alpha )=\frac{n!}{\alpha ^{n+1}}
\end{equation}

Let us briefly discuss the explicit construction of the trial wave functions which are used to approximate the exact wave functions of bound
states in the four-electron Be-atom. In this work we shall use the CI wave functions constructed from STO and $LS$
eigenfunctions. These wave functions are represented in the form
\begin{equation}
  \Psi = \sum_{p=1}^NC_p\Phi _p,\qquad \Phi _p = \hat{O}(\hat{L}^2) \hat{\mathcal{A}} \phi _p\chi
\end{equation}
i.e. it is a linear combination of $N$ symmetry adapted configurations $\Phi _p$, where the coefficients $C_p$  are determined variationally
by solving the eigenvalue problem which follows from the Schr\"{o}dinger equation.

In this work, the symmetry adapted configurations are
constructed 'a priori' so that they are eigenfunctions of the angular
momentum operator $\hat{L}^2$. Another possibility would be the posterior
projection of the configurations over the proper spatial space, as
indicated in Eq. (12) by the projection operator $\hat{O}(\hat{L}^2)$, where $\hat{\mathcal{A}%
}$ is the anti-symmetrization operator and $\chi$ is the spin eigenfunction for $S=0$ and $M_S=0$.
\begin{equation}
\chi =\left[ (\alpha \beta -\beta \alpha )(\alpha \beta -\beta \alpha
)\right] \
\end{equation}
As discussed in the case of the Li-atom in Ref. \cite{Li} and calculations of the Be-atom \cite{SH2011} it is sufficient to consider only one
spin-function. Strictly, a linear combination of all possible spin eigenfunctions should be employed but that it has proved not to be necessary.
Indeed, the Slater determinants produced by the anti-symmetrization of further spin
functions would be repeated when considering the spin eigenfunction Eq. (13).
The spatial part of the basis functions consists of Hartree products of STOs:
\begin{equation}
\phi _p=\prod_{k=1}^n\phi _k(r_k,\theta _k,\varphi _k).
\end{equation}
The basis functions $\phi _p$, are products of $s$-, $p$-, $d$-, $f$-, $g$-,
$h$-, $i$- and $k$-STOs, defined as

\begin{equation}
\phi (\mathbf{r})=r^{n-1}e^{-\alpha r}Y_l^m(\theta ,\varphi )
\end{equation}
where $Y_l^m(\theta ,\varphi )$ are the spherical harmonics \cite{Stevenson}.

We have written a four-electron CI computer program for four-electron atomic systems in Fortran 90. Numerical calculations have been conducted in double precision arithmetic. This program has been thoroughly checked by comparing results of our numerical calculations with the results by Sims
and Hagstrom for the Be atom \cite{SH71,SH2011}. In these calculations, we obtain complete agreement.

The ground state configuration of the Be-atom is $ssss$ (i.e. $%
s(1)s(2)s(3)s(4)$). The other configurations considered for S-symmetry
states (L=0) are, ordered by decreasing energy contribution, $sspp$, $spps
$, $ppss$, $pppp$, $ssdd$, $sdds$, $ddss$, $sppd$, $dpps$, $\ldots$, $ffdd$. The
energetically important configurations for $L=1,\cdots ,7$ are listed in
Table I. The quantum number $M_L=0$ was chosen, because for this case a smaller
number of Slater determinants is required.
We systematically selected the CI
configurations according to their energy contribution. This was done by
calculations on blocks constructed for all possible configurations.
The eigenvalue equation was diagonalized upon each addition of a configuration. In this manner, the contribution
of every single configuration and of each block of a given type to
the total energy was evaluated. Configurations with an overall energy contribution below $1\cdot 10^{-8}$ a.u.
were neglected.

The procedure of selection of the configurations is similar to that
described in our previous work, Refs. \cite{Our1,Li}. In this work we construct the full-CI (FCI) wave function
for every symmetry and basis set including the types of
configurations which contribute most to the total energy of the lowest state
of every symmetry. The larger the contribution of a configuration, the
smaller the sum of the $l$ quantum numbers of the employed orbitals $%
l_1+l_2+l_3+l_4$ is; i.e. the contribution of the configuration $sssp>sppp$
for a $P$-state. In cases such as the P states $sspd$ and $ppsp$, where the
sum of $l_i$ is equal, the two inner electrons in $ppsp$ form an
$S$-configuration. The resulting four-electron configuration is $(^1S)sp$ (a
$P$-configuration), and contributes more than the $sspd$ configuration.
This is especially important in the case of
$F$-, $G$-, $H$-, $I$- and $K$-states. Among the many possibilities to construct configurations of these symmetries,
the energetically most important configurations were proven to be those with an inner $S$-shell and a single occupied
orbital with the symmetry of the state under consideration, i.e. $(^1S)sf$, $(^1S)sg$, $(^1S)sh$, $(^1S)si$ and $(^1S)sk$. The inner shell is described with a
sum of configurations $(^1S)= ss+pp+dd+ff+gg+hh+ii+kk$. In the CI calculations of $S$-, $P$-, and $D$ states we used $s$-, $p$-, $d$-, and $f$-orbitals (see Table I).
In those of the $F$-, $G$-, $H$-, $I$- and $K$-states we have also used $g$-, $h$-, $i$- and $k$-orbitals (Table I).

More types of configurations than the ones discussed here can be constructed
for a given $L$ quantum number. For instance, configurations like $pssp$ could
be considered, if the exponents $\alpha _1\ne \alpha _2$. However, we have
kept the orbital exponents in the K-shell and L-shell equal.

Note that there are more possible 'degenerate L-eigenfunction' solutions with a larger number of
Slater determinants. Specifically, these are degenerate with respect to the quantum numbers L and M,
but with possible different energy contribution, i.e. non-degenerate with
respect to the energy \cite{Bunge}. Although the
inclusion of various degenerate configurations has been shown to improve
the energy of the state, such a contribution is very small. This is important
for very accurate CI calculations, as reported e.g. by Bunge
\cite{Bunge}. In our work, we have concentrated on the energetically most significant CI configurations.

Another important aspect in CI and Hy-CI calculations is the symmetry
adaptation of the configurations. As mentioned above, the configurations are constructed 'a
priori' to be eigenfunctions of the angular momentum operator $%
\hat{L}^2$. The configurations of Table I are constructed as sums of Slater
determinants. The determinants are pairwise symmetric (i.e. $ssp_1p_{-1}$ and
$ssp_{-1}p_1$ in the $sspp$ configuration) and lead to the same values of the
electronic integrals. Therefore, it is possible and desirable to consider
only one of the determinants and to deduce the second result.
In other words, the solution of the eigenvalue problem obtained when using
reduced $1\times 1$ matrix
elements (where the integrals are added, configuration $%
sp_1p_{-1}+sp_{-1}p_1$) or when using explicit $2\times 2$ matrix elements of the
Slater determinants is the same. The symmetry adaption is
computationally favorable, since the number of Slater determinants in the input is
smaller and the repeated computation of equal integrals is avoided.
The explicit sums of symmetry adapted configurations in the three-electron case are listed
in \cite{Li}.

In this work we start with the full-CI wave function (FCI) constructed with configurations of the type
of the ground configuration of a given state (see first configuration of every symmetry in Table I) and we use
the large basis set $n=8$. The notation $n=4$ stands for the basis set $[4s3p2d1f]$.
The first step consists in an optimization of the orbital exponents for this truncated wave function.
The optimization is carried out using a parabolic procedure, explained in Ref.
\cite{Li}. The orbital exponents are optimized for each atomic state of the
Be-atom. A set of two exponents is used (one for the $K$-shell and the other
for the electrons in the $L$-shell), and kept equal for all configurations.
This technique accelerates computations, while still producing
sufficiently accurate wave functions to determine the bound state properties. We use the virial theorem:
\begin{equation}
\chi =-\frac{\langle V\rangle }{\langle T\rangle }
\end{equation}
as a criterion to check the quality of the wave function and guide the numerical optimization of the exponents in the trial wave functions.

Using the appropriate exponents for every state we filtered the configurations of the first configuration block of the
FCI wave function calculating the total energy $E_i$ each time a single configuration was added,
and comparing it to the total energy without this configuration $E_{i-1}$.
If the difference of the energy was smaller than the
threshold $|E_{i-1}-E_i|<1\cdot 10^{-7}$a.u, the new configuration was discarded. In
this manner, all configurations were checked, leading to a relatively compact CI wave function.

The next step consists in adding a new block of configurations (FCI) of the following types given in Table I.
As the wave function may become very large, a new selection of the newly added configurations is carried out.
The resulting compact wave function is optimized again. The procedure is repeated when each new block of configurations is added.
In this work we employ the basis set $n=8$.
Using this method we obtain precise energy values and the addition of configurations with higher $l_i$ quantum numbers
contributes to the convergence to the non-relativistic energy.
The final wave function is a compact wave function containing one to two thousand
configurations which has milli-Hartree accuracy for the lowest states of every symmetry.
This technique is a compromise between selection and optimization.

By using the CI method we have calculated the bound $S$-, $P$-, $D$-, $F$-, $G$-, $H$-, $I$, and $K$-states in the Be-atom. In particular, we have
determined the energies of the four (lowest) S-states, three P-, three D-, two F-, two G-, one H-, on I-state and one K-state.
The total energies of the $F$-, $G$-, $H$-, $I$, and $K$-states in the Be atom are reported here for the first time.
To our knowledge, they have not been determined in earlier studies
neither computationally, nor experimentally. Our results are summarized in Table II. The overall accuracy of our calculations for the lowest states
of every symmetry can be evaluated as $\approx \pm 1-5 \cdot 10^{-3}$ $a.u.$ Higher excited states are not necessarily less accurate. The results have this limited accuracy due to the exponent restrictions of the method used here.
In Table III the CI calculations of the triplets bound states of the Be isoelectronic ions B$^+$, C$^{2+}$,
F$^{5+}$ amd Mg$^{8+}$ are summarized.

\section{General structure of the bound state spectra}

As mentioned above, the Be-atom bound state spectrum contains two series of bound states: singlet series and triplet series. To the lowest
order non-relativistic dipole approximation, these two series are independent of each other, i.e. any dipole transition between the two states
from different spectral series is strictly prohibited. In reality, transitions between singlet and triplet states of the Be-atom and other Be-like ions
are always possible due to non-elastic collisions of these atoms with electrons, ions and other atoms. It is clear that the probabilities of
such collisional transitions substantially depend upon the spatial densities of electrons, Be-atoms, etc. In very good vacuum ($\approx 10^{-12}$ $atm$)
and at relatively large temperatures one can easily see the two different optical series (singlet and triplet) in the gaseous mixture of the ${}^{9}$Be
atoms. Note also that very small relativistic components of the exact four-electron wave functions also make these singlet-triplet transitions possible.
Rates of such transitions are very low for the neutral Be-atom but they rapidly increase with the nuclear charge in the series: B$^+$, C$^{2+}$,
F$^{5+}$, Mg$^{8+}$. The electronic structure of the Be-atom ground singlet state is $1s^2 2s^2$, while all excited states have a similar structure
where the two electrons occupy the $1s^2$-electron shell (its excitation energy is extremely large), while the third electron is mostly located in the $2s$-shell.
% this small part of the text seems to be repeated:
%
%The electronic structure of the ground singlet state of the Be atom is $1s^2 2s^2$, while all excited states have a similar structure where
%the two electrons occupy the $1s^2$-electron shell (its excitation energy is extremely large), while the third electron is always located in
%the $2s$-shell.

The fourth (and sometimes the third) electron/s can occupy any free electron orbital in the atom. The occupation numbers of the
(third, fourth) electron/s
determine the actual state (or configuration) of the Be-atom. The pair of the third and fourth electron can be either in the singlet state,
or in the triplet state (as the whole Be-atom). It follows from here that the bound state spectrum of the Be-atom must be similar to the
bound state spectra of the two-electron He-atom. Indeed, such a similarity can be observed (the two series of bound states, the ground state is
the singlet $S$-state). However, the actual order of different bound states is different for the He- and Be-atoms. For instance, the lowest
state in the triplet series is $2^{3}S$-state in the helium atom and $2^{3}P$-state in the beryllium atom. For the excited bound states in the He-
and Be-atoms one finds more differences than similarities, while for singlet states close to the ground state similarities with between the spectra
of these two elements can easily be seen.

Since the three-electron core of the Be-atom has the $1s^2 2s$ electron configuration, then the dissociation threshold for neutral Be
corresponds to formation of the three-electron Be$^{+}$ ion in its ground $2^2S$-state (doublet). The non-relativistic energy of this state is
$E_{{}^{\infty}{\rm Be}^{+}}$ $\approx$ -14.324 763 176 790 43(22) $a.u.$ \cite{PKP}. This dissociation threshold corresponds to the following
ionization process of the Be atom
\begin{eqnarray}
  {\rm Be} = {\rm Be}^{+}(2^2S) + e^{-} \label{Rydb0}
\end{eqnarray}
where the symbol Be$^{+}(2^2S)$ means that the final three-electron Be$^{+}$ ion is in its ground $2^2S$-state. Now we can write the following
expression for the total energies of the weakly-bound states, i.e. for the states close to the dissociation threshold of the Be-atom
(in atomic units):
\begin{eqnarray}
  E({\rm Be}; n L) = E({\rm Be}^{+}; 2^2S) - \frac{m_e e^4}{2 \hbar^2} \frac{1}{(n + \Delta_{\ell})^2} =
  -14.324 763 176 790 43 - \frac{1}{2 (n + \Delta_{\ell})^2} \label{Rydb}
\end{eqnarray}
where $L = \ell$ (in this case), $n$ is the principal quantum number of the $n L$ state ($L$ is the angular quantum number) of the Be-atom and
$\Delta_{\ell}$ is the Rydberg correction which explicitly depends upon $\ell$ (angular momentum of the outer most electron) and the total
electron spin of this atomic state. It can be shown that the Rydberg correction rapidly vanishes when $\ell$ increases (for given $n$ and $L$).
Moreover, the $\Delta_{\ell}$ correction also decreases when the principal quantum number $n$ grows. The formula, Eq.(\ref{Rydb}), can be used to
approximate the total energies of weakly bound, Rydberg states in the Be-atom. However, by following the original ideas of Heisenberg \cite{Heis}
and Bethe (see, e.g., \cite{BS} and references therein) we can write a significantly more accurate formula which can be used to approximate the
same Rydberg states to very high numerical accuracy. This formula is written in the form
\begin{eqnarray}
  E({\rm Be}; n L) &=& E({\rm Be}^{+}; 2^2S) - \frac{m_e e^4}{2 \hbar^2} \frac{1 - \epsilon_{L}}{(n + \Delta_{\ell} + (-1)^{S} \Delta_{A})^2}
 \nonumber \\
                   &\approx& -14.324 763 176 790 43 - \frac{1 - \epsilon_L}{2 (n + \Delta_{\ell} + (-1)^{S} \Delta_{A})^2} \label{Rydb2}
\end{eqnarray}
where $S$ is the total electron spin, while $\epsilon_L, \Delta_{\ell}$ and $\Delta_A$ are the three parameters which must be varied in each specific case
to obtain better numerical approximations. All these parameters rapidly decrease when $\ell$ (and $L$) grows. In reality, to apply the formula,
Eq.(\ref{Rydb2}), one needs to know the accurate values of the total energies of at least three bound states in each spectral series, i.e. the total
energies of three singlet and three triplet bound states with $n \ge 5$.

Based on Eq.(\ref{Rydb2}) one can predict that the total energies of the singlet and triplet highly excited states (with the same $n$) are equal to
each other to high accuracy (near degeneracy). In general, such near degeneracy of energy levels becomes almost exact when $n$ grows. It is a well
known property of the Rydberg states and it can be observed in any atomic system which has energy spectrum consisting of a few different spectral
series. Formally, based on the formula, Eq.(\ref{Rydb2}), we can classify all bound states in the Be atom as the Rydberg states, pre-Rydberg and
non-Rydberg states. Each group of these states has its unique electron density distribution.

\section{Spectral diagram of the four-electron Be-like atoms}

In this study we have determined the total energies of a large number of bound singlet states in the Be-atom. Our computational results can be used
to draw the energy levels of all computed singlet (bound) states of the ${}^{\infty}$Be atom as functions of angular momentum $L$ of these states.
In classic books on atomic spectroscopy such pictures (or diagrams) were called the `spectral diagrams'. In general, the spectral diagrams are very useful
tools to study various effects related to the electron density distribution in different bound $LS$-states of the atomic systems which contain the
same number of electrons.

For neutral atoms and ions with the same nuclear charge $Q$, measured spectral diagrams are often used to investigate effects related
with the role of electron-electron correlations in different atomic states. For instance, from our spectral diagram one finds that the $3^{1}D$-state in
the Be-atom is less bound than the analogous $3^{1}S$-state, while for the bound $4^{1}D$- and $4^{1}S$-states such an order of bound states is reversed.
The true theory of electron-electron correlation in atoms must explain the observed order of the bound states (or energy levels) in the spectrum and
approximately predict the energy differences between them.

In general, by performing numerical calculations of a large number of bound states in atomic systems one always needs to answer the following two questions:
(1) predict the correct order of low-lying bound states, and (2) describe transitions between the low-lying bound states and weakly-bound, or Rydberg states.
To solve the first problem we can compare our results with the known experimental data for Be-atom \cite{Kramida}. For the singlet states in the Be-atom the
agreement between our computational results, the picture Fig.1 and data for the beryllium atom presented in \cite{Kramida} can be considered as very
good. Combining the theoretical and experimental data we can predict the total order of states in the singlet and triplet series, as shown in Table IV.
It is also clear we have calculated only the non-relativistic (total) energies, i.e. all relativistic and lowest-order QED corrections were ignored.
Note also that the CI method using STOs is substantially more accurate (in the order of few milli-Hartree) than various procedures based on Hartree-Fock 
approximation, but it still provides a restricted
description of the electron-electron correlation in specific atoms and ions. Nevertheless, the observed agreement with the measured bound state spectra of the
triplet states in the Be atom (or ${}^{9}$Be atom) is very good not only for low-lying bound states, but also for Rydberg states.

Now, consider the second problem. As follows from the results of our calculations all bound singlet states with $n \ge 6$ in the beryllium atom are the
weakly-bound, or Rydberg states. On the other hand, all bound triplet states in the Be-atom with $n \ge 4$ can be considered as the pre-Rydberg states. It
follows from comparison of the total energies of the triplet $4^{3}F$, $5^{3}G$ states and singlet $4^{1}F$, $5^{1}G$ states (for more details, see
\cite{Our1}). On the other hand, it is clear that the `boundary' principal quantum number $n_R$ from which the Rydberg states begin (for $n \ge n_R$) must
be exactly the same for both spectral series in the four-electron atoms and ions.

In this study our main focus was on the singlet bound states in the four-electron Be-atom. The triplet states in the Be-atom were considered in
our paper \cite{Our1} which also contains the spectral diagram of the bound triplet states in the Be-atom. For maximal convenience of the reader we also
included the updated spectral diagram of the triplet states in our present analysis (see, Fig.2). In many cases it is useful to observe spectral diagrams for
the both singlet and triplet series together and compare these diagrams with the analogous spectral diagrams for atoms which have bound state spectra
represented as a combination of the two separate spectral series (singlet and triplet). For instance it is very interesting and informative to make such a
comparison of the spectra of the beryllium and helium atoms (see, e.g., \cite{BS} and references therein).

We have also drawn spectral diagrams of the triplet states for various positively charged ions, e.g.,
for the Be atom and B$^{+}$, C$^{2+}$, F$^{5+}$, Mg$^{8+}$, see Figs. 3-6 using the computational
results of this work and the known experimental data from NIST Atomic Database and Refs. \cite{Nist-B,Nist-BCN,Nist-F,Nist-Mg}. The spectra of the negatively charged ions, e.g., the Li$^{-}$
ion) contains only a very few bound states (usually one bound state \cite{Fro99}) and its spectral diagram is very simple and not
informative. The corresponding spectral diagrams of the cations are very similar to each other, but there are few differences between them due to the
$Z$-dependence. These differences may well improve our understanding of the electron-electron correlation in the four-electron atomic systems.

\section{Conclusion}

We have considered the bound state spectrum of the singlet states in the four-electron Be-atom and the spectra of the triplet states the Be-like ions
B$^+$, C$^{2+}$, F$^{5+}$ and Mg$^{8+}$. The analogous spectrum of the bound triplet states in the
four-electron Be-atom has been presented and discussed in our earlier study \cite{Our1}. The results of both studies reproduce the complete `optical' spectrum
of the four-electron Be-atom. The agreement between our computational results and actual singlet/triplet spectrum of the Be-atom \cite{Kramida} can
be considered as very good. The quite complicated bound state spectrum of beryllium atom has been obtained and studied by using only computational methods.
This study is based on a computational approach which has three following advantages: (1) it can be applied for accurate computations of all bound states
in the spectrum, including rotationally excited states and weakly-bound, Rydberg atomic states, (2) it provides overall accuracy in the order of few milli-Hartree which is beyond the level provided
by various method based on the Hartree-Fock approximation, (3) such an accuracy does not decrease for the excited $LS$-states in the spectrum.
Disadvantages of this approach is the slow convergence of the CI method, which requires selection of predominant configurations and successive optimization
of the orbital exponents.

The results of this study allowed us to draw the spectral diagrams of the singlet and triplet spectra of the four-electron Be-atom and Be-like ions.
Such spectral diagrams for Be-atom can now be compared with analogous spectral diagrams of other light atoms and ions.
Theoretical comparison of the atomic spectra of the Be- and He-atoms seems to be very interesting, since each of these two spectra contains two independent
series of bound states (singlet and triplet).

\section{Acknowledgments}

We would like to thank James Sims and Stanley Hagstrom for providing us with numerical results from Hylleraas-CI calculations on the Be atom which have
serve to check our Be computer program, and to Carlos Bunge for valuable advise about the construction of $LS$ and degenerate configurations. One of us
(F.L.) greatly acknowledges his PhD supervisor Leticia Gonz\'alez at the University of Vienna for her scientific support. Finally, we are very grateful to 
Philip Hoggan and Telhat \"Ozdogan for the invitation to contribute to this Volume and for proofreading the manuscript.

\newpage

% PIC1  Be  ATOM Singlet states

\begin{sidewaysfigure}

\begin{center}
  \sansmath
  \begin{tikzpicture}[
    font=\sffamily,
    level/.style={black,thick},
    ionization/.style={black,dashed},scale=1.0
  ]
  \coordinate (sublevel) at (0, 8pt);

  \node at (-0.25,10.0) {n};
  \node at (0.5, 10.5){$^1$S};
  \node at (2.5, 10.5){$^1$P}; 
  \node at (4.5, 10.5){$^1$D};
  \node at (6.5, 10.5){$^1$F}; 
  \node at (8.5, 10.5){$^1$G};
  \node at (10.5, 10.5){$^1$H};
  \node at (12.5, 10.5){$^1$I}; 
  \node at (14.5, 10.5){$^1$K};
% \node at (16.5, 10.5){$^1$L};
% \node at (18.5, 10.5){$^1$M};
% \node at (20.5, 10.5){$^1$N};

  % S levels
  \node at (0.5,-0.20) {\scriptsize $2s^2$};
  \node at (0.5, 7.12) {\scriptsize $2s3s$};
  \node at (0.5, 8.62) {\scriptsize $2s4s$};
  \node at (0.5, 9.09) {\scriptsize $2s5s$};
% \node at (0.5, 9.34) {\scriptsize $2s6s$};
% \node at (0.5, 7.36) {\scriptsize $2s7s$};
% \node at (0.5, 7.91) {\scriptsize $2s8s$};

  \coordinate (S00) at (0, 0.0 );
  \coordinate (S10) at (0, 7.32);
  \coordinate (S20) at (0, 8.82);
  \coordinate (S30) at (0, 9.29);
  \coordinate (S40) at (0, 9.54);
% \coordinate (S50) at (0,  .  );
% \coordinate (S60) at (0,  .  );

  % Draw main levels
   \foreach \level/\text in { 00/2,  10/3,  20/4,  30/5, 40/6}
%  \foreach \level/\text in { 00/2,  10/3,  20/4,  30/5, 40/6, 50/7, 60/8}
     \draw[level] (S\level) node[left=20pt] {} node[left]
      {\scriptsize {\text}} -- +(1.0, 0);

  % P levels          

  % ^1P levels          
  \node at (2.5,5.53) {\scriptsize $2s2p$};
  \node at (2.5,7.91) {\scriptsize $2s3p$};
  \node at (2.5,8.95) {\scriptsize $2s4p$};
  \node at (2.5,9.50) {\scriptsize $2s5p$};

%  \node at (2.5,9.10) {\scriptsize $6p$};
%  \node at (2.5,9.11) {\scriptsize $7p$};
%  \node at (2.5,9.42) {\scriptsize $8p$};

  \coordinate (P00) at (2, 5.73);
  \coordinate (P10) at (2, 8.11);
  \coordinate (P20) at (2, 9.15);
  \coordinate (P30) at (2, 9.70);
% \coordinate (P40) at (2, 9.54);
% \coordinate (P50) at (2, 9.67);

  % Draw main levels
%  \foreach \level/\text in {00/2, 10/3, 20/4, 30/5, 40/6, 50/7}
   \foreach \level/\text in {00/2, 10/3, 20/4, 30/5}
    \draw[level] (P\level) node[left=20pt] {} node[left]
     {\footnotesize {\text}} -- +(1.0, 0);

  % ^1D levels          
  \node at (4.5,7.58) {\scriptsize $2p^2$};
  \node at (4.5,8.56) {\scriptsize $2s3d$};
  \node at (4.5,9.25) {\scriptsize $2s4d$};
  \coordinate (D00) at (4, 7.78);
  \coordinate (D10) at (4, 8.76);
  \coordinate (D20) at (4, 9.45);
%  \coordinate (D30) at (4, 9.44);
%  \coordinate (D40) at (4, 9.60);

  % Draw main levels
%  \foreach \level/\text in {00/3, 10/4, 20/5, 30/6, 40/7}
  \foreach \level/\text in {00/3, 10/4, 20/5}
    \draw[level] (D\level) node[left=20pt] {} node[left]
     {\footnotesize {\text}} -- +(1.0, 0);

% ^1F levels      
  \node at (6.5,8.98) {\scriptsize $2s4f$};
  \node at (6.5,9.51) {\scriptsize $2s5f$};
  \coordinate (F00) at (6, 9.18);
  \coordinate (F10) at (6, 9.71);

  % Draw main levels
  \foreach \level/\text in {00/4, 10/5}
    \draw[level] (F\level) node[left=20pt] {} node[left]
     {\footnotesize {\text}} -- +(1.0, 0);

  % ^1G levels          
  \node at (8.5,9.40) {\scriptsize $2s5g$};
  \coordinate (G00) at (8,9.60);

  % Draw main levels
  \foreach \level/\text in {00/5}
    \draw[level] (G\level) node[left=20pt] {} node[left]
     {\footnotesize {\text}} -- +(1.0, 0);

  % ^1H levels          
  \node at (10.5,9.50) {\scriptsize $2s6h$};
  \coordinate (H00) at (10,9.70);

  % Draw main levels
  \foreach \level/\text in {00/6}
    \draw[level] (H\level) node[left=20pt] {} node[left]
     {\footnotesize {\text}} -- +(1.0, 0);

  %   I levels 
  \node at (12.5,9.50) {\scriptsize $2s7i$};
%  \node at (12.5,9.28) {\scriptsize $8i$};
  \coordinate (I00) at (12,9.70);

  % Draw main levels
  \foreach \level/\text in {00/7}
    \draw[level] (I\level) node[left=20pt] {} node[left]
    {\footnotesize {\text}} -- +(1.0, 0);

  % K levels

  \node at (14.5,9.50){\scriptsize $2s8k$};
  \coordinate (K00) at (14,9.70);

  % Draw main levels
  \foreach \level/\text in {00/8}
    \draw[level] (K\level) node[left=20pt] {} node[left]
      {\footnotesize {\text}} -- +(1.0, 0);

  % Ionization level
  \draw[ionization] (0, 10.0) node[left=20pt] {E(${\rm Be}^+$)}-- +( 15.0, 0);

  % Rydberg levels
  \draw[level] (0,9.8) node[left=20pt] {}-- +(1.0, 0);
  \draw[level] (0,9.9) node[left=20pt] {}-- +(1.0, 0); 

  \draw[level] (2,9.8) node[left=20pt] {}-- +(1.0, 0);
  \draw[level] (2,9.9) node[left=20pt] {}-- +(1.0, 0);

  \draw[level] (4,9.8) node[left=20pt] {}-- +(1.0, 0);
  \draw[level] (4,9.9) node[left=20pt] {}-- +(1.0, 0);

  \draw[level] (6,9.8) node[left=20pt] {}-- +(1.0, 0);
  \draw[level] (6,9.9) node[left=20pt] {}-- +(1.0, 0);

  \draw[level] (8,9.8) node[left=20pt] {}-- +(1.0, 0);
  \draw[level] (8,9.9) node[left=20pt] {}-- +(1.0, 0);

  \draw[level] (10,9.8) node[left=20pt] {}-- +(1.0, 0);
  \draw[level] (10,9.9) node[left=20pt] {}-- +(1.0, 0);

  \draw[level] (12,9.8) node[left=20pt] {}-- +(1.0, 0);
  \draw[level] (12,9.9) node[left=20pt] {}-- +(1.0, 0);

  \draw[level] (14,9.8) node[left=20pt] {}-- +(1.0, 0);
  \draw[level] (14,9.9) node[left=20pt] {}-- +(1.0, 0);

% \draw[level] (16,9.8) node[left=20pt] {}-- +(1.0, 0);
% \draw[level] (16,9.9) node[left=20pt] {}-- +(1.0, 0);

% \draw[level] (18,9.8) node[left=20pt] {}-- +(1.0, 0);
% \draw[level] (18,9.9) node[left=20pt] {}-- +(1.0, 0);

% \draw[level] (20,9.8) node[left=20pt] {}-- +(1.0, 0);
% \draw[level] (20,9.9) node[left=20pt] {}-- +(1.0, 0);

  \end{tikzpicture}
\end{center}

\vspace{1cm}

\caption{The energy levels of the singlet states in the beryllium atom.  
The threshold energy (or ionization limit) $E_{{}^{\infty}{\rm Be}^{+}}$ = -14.324 763 176 790 43(22) $a.u.$ coincides with the total energy of the
ground 2$^2S$-state of the three-electron Be$^{+}$ ion.}

\end{sidewaysfigure}

\newpage

\begin{sidewaysfigure}

% PIC2     TRIPLET STATES (small) 

\begin{center}
  \sansmath
  \begin{tikzpicture}[
    font=\sffamily,
    level/.style={black,thick},
    ionization/.style={black,dashed},
  ]
  \coordinate (sublevel) at (0, 8pt);

  \node at (0.5, 10.5) {$^3$S};
  \node at (2.5, 10.5) {$^3$P}; 
  \node at (4.5, 10.5) {$^3$D};
  \node at (6.5, 10.5) {$^3$F}; 
  \node at (8.5, 10.5) {$^3$G};
  \node at (10.5, 10.5) {$^3$H};
  \node at (12.5, 10.5) {$^3$I};
  \node at (14.5, 10.5) {$^3$K};
  \node at (-0.25,10.0) {n};

  % S levels
  \node at (0.5, 5.46) {\scriptsize $2s3s$};
  \node at (0.5, 7.40) {\scriptsize $2s4s$};
  \node at (0.5, 8.70) {\scriptsize $2s5s$};
  \node at (0.5, 9.50) {\scriptsize $2s6s$};
  \coordinate (S00) at (0, 5.66);
  \coordinate (S10) at (0, 7.6);
  \coordinate (S20) at (0, 8.90);
  \coordinate (S30) at (0, 9.70);

  % Draw main levels
   \foreach \level/\text in { 00/2,  10/3,  20/4,  30/5}
     \draw[level] (S\level) node[left=20pt] {} node[left]  
        {\footnotesize {\text}} -- +(1.0, 0);

  % P levels
  \node at (2.5,-0.20) {\scriptsize $2s2p$};      
  \node at (2.5, 6.74) {\scriptsize $2s3p$};     
  \node at (2.5, 8.25) {\scriptsize $2s4p$};
  \node at (2.5, 9.30) {\scriptsize $2s5p$};
  \coordinate (P00) at (2, 0.0);
  \coordinate (P10) at (2, 6.94);
  \coordinate (P20) at (2, 8.45);
  \coordinate (P30) at (2, 9.5);

  % Draw main levels
  \foreach \level/\text in {00/2, 10/3, 20/4, 30/5}
    \draw[level] (P\level) node[left=20pt] {} node[left] 
     {\footnotesize {\text}} -- +(1.0, 0);

  % D levels          
  \node at (4.5,7.36) {\scriptsize $2s3d$};
  \node at (4.5,8.40) {\scriptsize $2s4d$};
  \node at (4.5,9.35) {\scriptsize $2s5d$};
  \coordinate (D00) at (4, 7.56);
  \coordinate (D10) at (4, 8.6);
  \coordinate (D20) at (4, 9.55);

  % Draw main levels
  \foreach \level/\text in {00/3, 10/4, 20/5}
    \draw[level] (D\level) node[left=20pt] {} node[left]
     {\footnotesize {\text}} -- +(1.0, 0);

  % F levels          
  \node at (6.5,8.60) {\scriptsize $2s4f$};
  \node at (6.5,9.40) {\scriptsize $2s5f$};
  \coordinate (F00) at (6, 8.80);
  \coordinate (F10) at (6, 9.60);

  % Draw main levels
  \foreach \level/\text in {00/4, 10/5 }
    \draw[level] (F\level) node[left=20pt] {} node[left]
     {\footnotesize {\text}} -- +(1.0, 0);

  % G levels       
  \node at (8.5,9.45) {\scriptsize $2s5g$};   
  \coordinate (G00) at (8,9.65);

  % Draw main levels
  \foreach \level/\text in {00/5}
    \draw[level] (G\level) node[left=20pt] {} node[left]
     {\footnotesize {\text}} -- +(1.0, 0);

  % Ionization level
  \draw[ionization] (0, 10.0) node[left=20pt] {E(${\rm Be}^+$)}-- +( 15, 0);

  % Rydberg levels
  \draw[level] (0,9.8) node[left=20pt] {}-- +(1.0, 0);
  \draw[level] (0,9.9) node[left=20pt] {}-- +(1.0, 0); 

  \draw[level] (2,9.8) node[left=20pt] {}-- +(1.0, 0);
  \draw[level] (2,9.9) node[left=20pt] {}-- +(1.0, 0);

  \draw[level] (4,9.8) node[left=20pt] {}-- +(1.0, 0);
  \draw[level] (4,9.9) node[left=20pt] {}-- +(1.0, 0);

  \draw[level] (6,9.8) node[left=20pt] {}-- +(1.0, 0);
  \draw[level] (6,9.9) node[left=20pt] {}-- +(1.0, 0);

  \draw[level] (8,9.8) node[left=20pt] {}-- +(1.0, 0);
  \draw[level] (8,9.9) node[left=20pt] {}-- +(1.0, 0);

  \draw[level] (10,9.8) node[left=20pt] {}-- +(1.0, 0);
  \draw[level] (10,9.9) node[left=20pt] {}-- +(1.0, 0);

  \draw[level] (12,9.9) node[left=20pt] {}-- +(1.0, 0);

  \draw[level] (14,9.9) node[left=20pt] {}-- +(1.0, 0);

  \node at (10.5,9.60) {\scriptsize $2s6h$};
  \node at (12.5,9.70) {\scriptsize $2s7i$};
  \node at (14.5,9.70) {\scriptsize $2s8k$};

  \end{tikzpicture}

 \vspace{1cm}

\end{center}

\caption{The energy levels of the triplet states in the beryllium atom. The threshold
energy (or ionization limit) $E_{{}^{\infty}{\rm Be}^{+}}$ = -14.324 763 176 790 43(22) $a.u.$ coincides with the total energy of the
ground 2$^2S$-state of the three-electron Be$^{+}$ ion.}

\end{sidewaysfigure}

\newpage

% PIC1  B+ ION triplet states

\begin{sidewaysfigure}

\begin{center}
  \sansmath
  \begin{tikzpicture}[
    font=\sffamily,
    level/.style={black,thick},
    ionization/.style={black,dashed},scale=1.0
  ]
  \coordinate (sublevel) at (0, 8pt);

  \node at (-0.25,10.0) {n};
  \node at (0.5, 10.5){$^3$S};
  \node at (2.5, 10.5){$^3$P}; 
  \node at (4.5, 10.5){$^3$D};
  \node at (6.5, 10.5){$^3$F}; 
  \node at (8.5, 10.5){$^3$G};
  \node at (10.5, 10.5){$^3$H};
  \node at (12.5, 10.5){$^3$I}; 
  \node at (14.5, 10.5){$^3$K};

  % S levels
  \node at (0.5, 5.39) {\scriptsize $2s3s$};
  \node at (0.5, 7.61) {\scriptsize $2s4s$};
  \node at (0.5, 8.50) {\scriptsize $2s5s$};
  \node at (0.5, 9.05) {\scriptsize $2s6s$};
% \node at (0.5, 9.63) {\scriptsize $2s7s$};
% \node at (0.5, 9.36) {\scriptsize $2s8s$};

  \coordinate (S00) at (0, 5.59);
  \coordinate (S10) at (0, 7.81);
  \coordinate (S20) at (0, 8.70);
  \coordinate (S30) at (0, 9.25);
  \coordinate (S40) at (0, 9.57);
% \coordinate (S50) at (0, 9.83);
% \coordinate (S60) at (0, 9.56);

  % Draw main levels
   \foreach \level/\text in { 00/2,  10/3,  20/4,  30/5, 40/6}        
%  \foreach \level/\text in { 00/2,  10/3,  20/4,  30/5, 40/6, 50/7, 60/8}
     \draw[level] (S\level) node[left=20pt] {} node[left]
      {\footnotesize {\text}} -- +(1.0, 0);

  % P levels          

  % ^1P levels          
  \node at (2.5,-0.20) {\scriptsize $2s2p$};
  \node at (2.5,6.30) {\scriptsize $2p^2$};
  \node at (2.5,8.00) {\scriptsize $2s3p$};
  \node at (2.5,8.60) {\scriptsize $2s4p$};
  \node at (2.5,9.20) {\scriptsize $2p3s$};
% \node at (2.5,9.11) {\scriptsize $7p$};
%  \node at (2.5,9.42) {\scriptsize $8p$};

  \coordinate (P00) at (2, 0.0 );
  \coordinate (P10) at (2, 6.50);
  \coordinate (P20) at (2, 8.20);
  \coordinate (P30) at (2, 8.80);
  \coordinate (P40) at (2, 9.40);
%  \coordinate (P50) at (2, 9.95);

  % Draw main levels
   \foreach \level/\text in {00/2, 10/3, 20/4, 30/5, 40/6}         
%  \foreach \level/\text in {00/2, 10/3, 20/4}
    \draw[level] (P\level) node[left=20pt] {} node[left]
     {\footnotesize {\text}} -- +(1.0, 0);

% % ^1D levels          
  \node at (4.5,6.78) {\scriptsize $2s3d$};
  \node at (4.5,8.16) {\scriptsize $2s4d$};
  \node at (4.5,8.88) {\scriptsize $2s5d$};
  \coordinate (D00) at (4, 6.98);
  \coordinate (D10) at (4, 8.36);
  \coordinate (D20) at (4, 9.08);
  \coordinate (D30) at (4, 9.45);
%  \coordinate (D40) at (4, 9.60);

% % Draw main levels
   \foreach \level/\text in {00/3, 10/4, 20/5, 30/6}
% \foreach \level/\text in {00/3, 10/4}
    \draw[level] (D\level) node[left=20pt] {} node[left]
     {\footnotesize {\text}} -- +(1.0, 0);

% ^1F levels      
  \node at (6.5,8.18) {\scriptsize $2s4f$};
  \node at (6.5,8.92) {\scriptsize $2s5f$};
% \node at (6.5,9.27) {\scriptsize $2s6f$}; 
  \coordinate (F00) at (6, 8.38);
  \coordinate (F10) at (6, 9.12);
  \coordinate (F20) at (6, 9.47);

  % Draw main levels
  \foreach \level/\text in {00/4, 10/5, 20/6}
    \draw[level] (F\level) node[left=20pt] {} node[left]
     {\footnotesize {\text}} -- +(1.0, 0);

  % ^1G levels          
  \node at (8.5,9.30) {\scriptsize $2s5g$};
  \coordinate (G00) at (8,9.50);

  % Draw main levels
  \foreach \level/\text in {00/5}
    \draw[level] (G\level) node[left=20pt] {} node[left]
     {\footnotesize {\text}} -- +(1.0, 0);

  % ^1H levels          
  \node at (10.5,9.33) {\scriptsize $2s6h$};
  \coordinate (H00) at (10,9.53);
 
  % Draw main levels
  \foreach \level/\text in {00/6}
    \draw[level] (H\level) node[left=20pt] {} node[left]
     {\footnotesize {\text}} -- +(1.0, 0);

%   ^1I levels 
  \node at (12.5,9.35) {\scriptsize $2s7i$};
%  \node at (12.5,9.28) {\scriptsize $8i$};
  \coordinate (I00) at (12,9.55);

  % Draw main levels
  \foreach \level/\text in {00/7}
    \draw[level] (I\level) node[left=20pt] {} node[left]
    {\footnotesize {\text}} -- +(1.0, 0);

  % K levels
 
  \node at (14.5,9.36){\scriptsize $2s8k$};
  \coordinate (K00) at (14,9.56);

  % Draw main levels
  \foreach \level/\text in {00/8}
    \draw[level] (K\level) node[left=20pt] {} node[left]
      {\footnotesize {\text}} -- +(1.0, 0);

  % Ionization level
  \draw[ionization] (0, 10.0) node[left=20pt] {E(${\rm B}^{2+}$)}-- +( 15.0, 0);

  % Rydberg levels
  \draw[level] (0,9.8) node[left=20pt] {}-- +(1.0, 0);
  \draw[level] (0,9.9) node[left=20pt] {}-- +(1.0, 0); 

  \draw[level] (2,9.8) node[left=20pt] {}-- +(1.0, 0);
  \draw[level] (2,9.9) node[left=20pt] {}-- +(1.0, 0);

  \draw[level] (4,9.8) node[left=20pt] {}-- +(1.0, 0);
  \draw[level] (4,9.9) node[left=20pt] {}-- +(1.0, 0);

  \draw[level] (6,9.8) node[left=20pt] {}-- +(1.0, 0);
  \draw[level] (6,9.9) node[left=20pt] {}-- +(1.0, 0);

  \draw[level] (8,9.8) node[left=20pt] {}-- +(1.0, 0);
  \draw[level] (8,9.9) node[left=20pt] {}-- +(1.0, 0);

  \draw[level] (10,9.8) node[left=20pt] {}-- +(1.0, 0);
  \draw[level] (10,9.9) node[left=20pt] {}-- +(1.0, 0);

  \draw[level] (12,9.8) node[left=20pt] {}-- +(1.0, 0);
  \draw[level] (12,9.9) node[left=20pt] {}-- +(1.0, 0);

  \draw[level] (14,9.8) node[left=20pt] {}-- +(1.0, 0);
  \draw[level] (14,9.9) node[left=20pt] {}-- +(1.0, 0);

  \end{tikzpicture}
\end{center}

\vspace{1cm}

\caption{The energy levels of the triplet states in the B$^+$ ion. 
The threshold
energy (or ionization limit) $E($B$^{2+})$ =-23.424 605 665 $a.u.$ coincides with the total energy of the
ground 2$^2S$-state of the three-electron B$^{2+}$ ion.}

\end{sidewaysfigure}

\newpage

% PIC1 C2+ ION triplet states

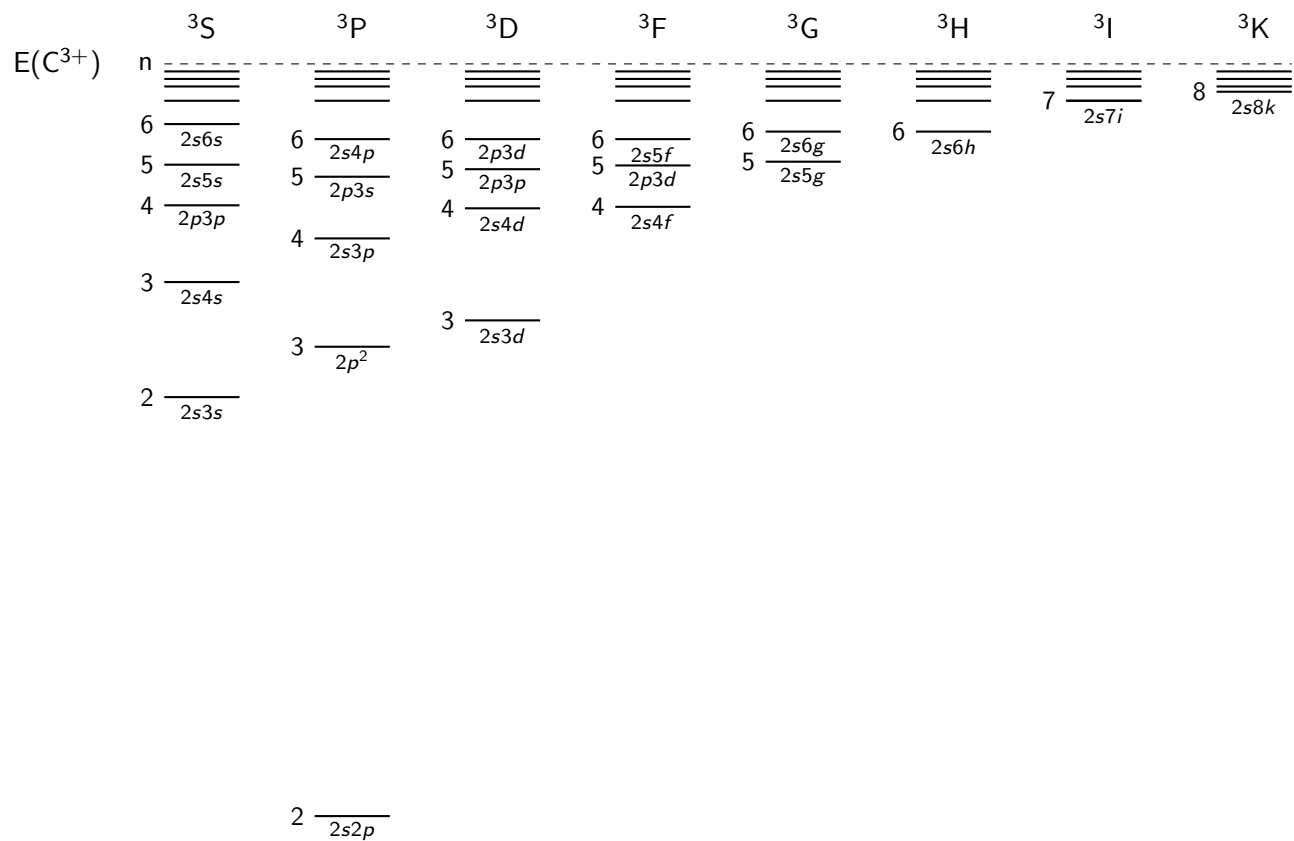
\begin{sidewaysfigure}

\begin{center}
  \sansmath
  \begin{tikzpicture}[
    font=\sffamily,
    level/.style={black,thick},
    ionization/.style={black,dashed},scale=1.0
  ]
  \coordinate (sublevel) at (0, 8pt);

  \node at (-0.25,10.0) {n};
  \node at (0.5, 10.5){$^3$S};
  \node at (2.5, 10.5){$^3$P}; 
  \node at (4.5, 10.5){$^3$D};
  \node at (6.5, 10.5){$^3$F}; 
  \node at (8.5, 10.5){$^3$G};
  \node at (10.5, 10.5){$^3$H};
  \node at (12.5, 10.5){$^3$I}; 
  \node at (14.5, 10.5){$^3$K};

  % S levels
  \node at (0.5, 5.37) {\scriptsize $2s3s$};
  \node at (0.5, 6.90) {\scriptsize $2s4s$};
  \node at (0.5, 7.92) {\scriptsize $2p3p$};
  \node at (0.5, 8.46) {\scriptsize $2s5s$};
  \node at (0.5, 9.00) {\scriptsize $2s6s$};
% \node at (0.5, 9.36) {\scriptsize $2s8s$};

  \coordinate (S00) at (0, 5.57);
  \coordinate (S10) at (0, 7.10);
  \coordinate (S20) at (0, 8.12);
  \coordinate (S30) at (0, 8.66);
  \coordinate (S40) at (0, 9.20);
% \coordinate (S60) at (0, 9.56);

  % Draw main levels
%  \foreach \level/\text in { 00/2,  10/3,  20/4,  30/5}
   \foreach \level/\text in { 00/2,  10/3,  20/4,  30/5, 40/6}
     \draw[level] (S\level) node[left=20pt] {} node[left]
      {\footnotesize {\text}} -- +(1.0, 0);

  % P levels          

  % ^1P levels          
  \node at (2.5,-0.20) {\scriptsize $2s2p$};
  \node at (2.5,6.04) {\scriptsize $2p^2$};
  \node at (2.5,7.48) {\scriptsize $2s3p$};
  \node at (2.5,8.30) {\scriptsize $2p3s$};
  \node at (2.5,8.80) {\scriptsize $2s4p$};
% \node at (2.5,8.70) {\scriptsize $2s7p$};

  \coordinate (P00) at (2, 0.0 );
  \coordinate (P10) at (2, 6.24);
  \coordinate (P20) at (2, 7.68);
  \coordinate (P30) at (2, 8.50);
  \coordinate (P40) at (2, 9.00);
% \coordinate (P50) at (2, 8.94);

  % Draw main levels
  \foreach \level/\text in {00/2, 10/3, 20/4, 30/5, 40/6}        
%   \foreach \level/\text in {00/2, 10/3, 20/4}
    \draw[level] (P\level) node[left=20pt] {} node[left]
     {\footnotesize {\text}} -- +(1.0, 0);

% % ^1D levels          
  \node at (4.5,6.39) {\scriptsize $2s3d$};
  \node at (4.5,7.88) {\scriptsize $2s4d$};
  \node at (4.5,8.40) {\scriptsize $2p3p$};
  \node at (4.5,8.80) {\scriptsize $2p3d$};
  \coordinate (D00) at (4, 6.59);
  \coordinate (D10) at (4, 8.08);
  \coordinate (D20) at (4, 8.60);
  \coordinate (D30) at (4, 9.00);
% \coordinate (D40) at (4, 9.60);

% % Draw main levels
   \foreach \level/\text in {00/3, 10/4, 20/5, 30/6}          
% \foreach \level/\text in {00/3, 10/4}
    \draw[level] (D\level) node[left=20pt] {} node[left]
     {\footnotesize {\text}} -- +(1.0, 0);

% ^1F levels      
  \node at (6.5,7.90) {\scriptsize $2s4f$};
  \node at (6.5,8.45) {\scriptsize $2p3d$};
  \node at (6.5,8.80) {\scriptsize $2s5f$};
  \coordinate (F00) at (6, 8.10);
  \coordinate (F10) at (6, 8.65);
  \coordinate (F20) at (6, 9.00);

  % Draw main levels
  \foreach \level/\text in {00/4, 10/5, 20/6}
    \draw[level] (F\level) node[left=20pt] {} node[left]
     {\footnotesize {\text}} -- +(1.0, 0);

  % ^1G levels          
  \node at (8.5,8.50) {\scriptsize $2s5g$};
  \node at (8.5,8.90) {\scriptsize $2s6g$};
  \coordinate (G00) at (8,8.70);
  \coordinate (G10) at (8,9.10);

  % Draw main levels
  \foreach \level/\text in {00/5, 10/6}
    \draw[level] (G\level) node[left=20pt] {} node[left]
     {\footnotesize {\text}} -- +(1.0, 0);

  % ^1H levels          
  \node at (10.5,8.90) {\scriptsize $2s6h$};
  \coordinate (H00) at (10,9.10);

  % Draw main levels
  \foreach \level/\text in {00/6}
    \draw[level] (H\level) node[left=20pt] {} node[left]
     {\footnotesize {\text}} -- +(1.0, 0);

%   ^1I levels 
  \node at (12.5,9.31) {\scriptsize $2s7i$};
%  \node at (12.5,9.28) {\scriptsize $8i$};
  \coordinate (I00) at (12,9.51);

  % Draw main levels
  \foreach \level/\text in {00/7}
    \draw[level] (I\level) node[left=20pt] {} node[left]
    {\footnotesize {\text}} -- +(1.0, 0);

  % K levels

  \node at (14.5,9.43){\scriptsize $2s8k$};
  \coordinate (K00) at (14,9.63);

  % Draw main levels
  \foreach \level/\text in {00/8}
    \draw[level] (K\level) node[left=20pt] {} node[left]
      {\footnotesize {\text}} -- +(1.0, 0);

  % Ionization level
  \draw[ionization] (0, 10.0) node[left=20pt] {E(${\rm C}^{3+}$)}-- +( 15.0, 0);

  % Rydberg levels
  \draw[level] (0,9.51) node[left=20pt] {}-- +(1.0, 0);
  \draw[level] (0,9.7) node[left=20pt] {}-- +(1.0, 0);
  \draw[level] (0,9.8) node[left=20pt] {}-- +(1.0, 0);
  \draw[level] (0,9.9) node[left=20pt] {}-- +(1.0, 0); 

  \draw[level] (2,9.51) node[left=20pt] {}-- +(1.0, 0);
  \draw[level] (2,9.7) node[left=20pt] {}-- +(1.0, 0);
  \draw[level] (2,9.8) node[left=20pt] {}-- +(1.0, 0);
  \draw[level] (2,9.9) node[left=20pt] {}-- +(1.0, 0);

  \draw[level] (4,9.51) node[left=20pt] {}-- +(1.0, 0);
  \draw[level] (4,9.7) node[left=20pt] {}-- +(1.0, 0);
  \draw[level] (4,9.8) node[left=20pt] {}-- +(1.0, 0);
  \draw[level] (4,9.9) node[left=20pt] {}-- +(1.0, 0);

  \draw[level] (6,9.51) node[left=20pt] {}-- +(1.0, 0);
  \draw[level] (6,9.7) node[left=20pt] {}-- +(1.0, 0);
  \draw[level] (6,9.8) node[left=20pt] {}-- +(1.0, 0);
  \draw[level] (6,9.9) node[left=20pt] {}-- +(1.0, 0);

  \draw[level] (8,9.51) node[left=20pt] {}-- +(1.0, 0);
  \draw[level] (8,9.7) node[left=20pt] {}-- +(1.0, 0);
  \draw[level] (8,9.8) node[left=20pt] {}-- +(1.0, 0);
  \draw[level] (8,9.9) node[left=20pt] {}-- +(1.0, 0);

  \draw[level] (10,9.51) node[left=20pt] {}-- +(1.0, 0);
  \draw[level] (10,9.7) node[left=20pt] {}-- +(1.0, 0);
  \draw[level] (10,9.8) node[left=20pt] {}-- +(1.0, 0);
  \draw[level] (10,9.9) node[left=20pt] {}-- +(1.0, 0);

  \draw[level] (12,9.51) node[left=20pt] {}-- +(1.0, 0);
  \draw[level] (12,9.7) node[left=20pt] {}-- +(1.0, 0);
  \draw[level] (12,9.8) node[left=20pt] {}-- +(1.0, 0);
  \draw[level] (12,9.9) node[left=20pt] {}-- +(1.0, 0);

  \draw[level] (14,9.7) node[left=20pt] {}-- +(1.0, 0);
  \draw[level] (14,9.8) node[left=20pt] {}-- +(1.0, 0);
  \draw[level] (14,9.9) node[left=20pt] {}-- +(1.0, 0);

  \end{tikzpicture}
\end{center}

\vspace{1cm}

\caption{The energy levels of the triplet states in the C$^{2+}$ ion. 
The threshold
energy (or ionization limit) $E($C$^{3+})$ =-34.775 510 611 $a.u.$ coincides with the total energy of the
ground 2$^2S$-state of the three-electron C$^{3+}$ ion.} 

\end{sidewaysfigure}

\newpage

% PIC1 F5+ ION triplet states

\begin{sidewaysfigure}

\begin{center}
  \sansmath
  \begin{tikzpicture}[
    font=\sffamily,
    level/.style={black,thick},
    ionization/.style={black,dashed},scale=1.0
  ]
  \coordinate (sublevel) at (0, 8pt);

  \node at (-0.25,10.0) {n};
  \node at (0.5, 10.5){$^3$S};
  \node at (2.5, 10.5){$^3$P}; 
  \node at (4.5, 10.5){$^3$D};
  \node at (6.5, 10.5){$^3$F}; 
  \node at (8.5, 10.5){$^3$G};
  \node at (10.5, 10.5){$^3$H};
  \node at (12.5, 10.5){$^3$I}; 
  \node at (14.5, 10.5){$^3$K};

  % S levels
  \node at (0.5, 5.16) {\scriptsize $2s3s$};
  \node at (0.5, 6.15) {\scriptsize $2p3p$};
  \node at (0.5, 7.60) {\scriptsize $2s4s$};
  \node at (0.5, 8.43) {\scriptsize $2s5s$};
  \node at (0.5, 8.80) {\scriptsize $2p4p$};
  \node at (0.5, 9.20) {\scriptsize $2s6s$};
%  \node at (0.5, 8.59) {\scriptsize $2s8s$};

  \coordinate (S00) at (0, 5.36);
  \coordinate (S10) at (0, 6.35);
  \coordinate (S20) at (0, 7.80);
  \coordinate (S30) at (0, 8.63);
  \coordinate (S40) at (0, 9.00);
  \coordinate (S50) at (0, 9.40);
% \coordinate (S60) at (0, 8.79);

  % Draw main levels
%  \foreach \level/\text in { 00/2,  10/3,  20/4,  30/5}
   \foreach \level/\text in { 00/2,  10/3,  20/4,  30/5, 40/6, 50/7}
     \draw[level] (S\level) node[left=20pt] {} node[left]
      {\footnotesize {\text}} -- +(1.0, 0);

  % P levels          

  % ^1P levels          
  \node at (2.5,-0.20) {\scriptsize $2s2p$};
  \node at (2.5,5.73) {\scriptsize $2p^2$};
  \node at (2.5,7.00) {\scriptsize $2s3p$};
  \node at (2.5,7.99) {\scriptsize $2p3s$};
  \node at (2.5,8.60) {\scriptsize $2p3p$};
  \node at (2.5,9.10) {\scriptsize $2p3d$};
% \node at (2.5,9.42) {\scriptsize $8p$};

  \coordinate (P00) at (2, 0.0 );
  \coordinate (P10) at (2, 5.93);
  \coordinate (P20) at (2, 7.20);
  \coordinate (P30) at (2, 8.19);
  \coordinate (P40) at (2, 8.80);
  \coordinate (P50) at (2, 9.30);

  % Draw main levels
   \foreach \level/\text in {00/2, 10/3, 20/4, 30/5, 40/6, 50/7}        
%  \foreach \level/\text in {00/2, 10/3, 20/4}
    \draw[level] (P\level) node[left=20pt] {} node[left]
     {\footnotesize {\text}} -- +(1.0, 0);

% % ^1D levels          
  \node at (4.5,5.95) {\scriptsize $2s3d$};
  \node at (4.5,7.65) {\scriptsize $2p3p$};
  \node at (4.5,8.44) {\scriptsize $2p3d$};
  \node at (4.5,8.81) {\scriptsize $2s4d$};
  \node at (4.5,9.15) {\scriptsize $2s5d$};
  \coordinate (D00) at (4, 6.15);
  \coordinate (D10) at (4, 7.85);
  \coordinate (D20) at (4, 8.64);
  \coordinate (D30) at (4, 9.01);
  \coordinate (D40) at (4, 9.35);

% % Draw main levels
   \foreach \level/\text in {00/3, 10/4, 20/5, 30/6, 40/7}          
% \foreach \level/\text in {00/3, 10/4}
    \draw[level] (D\level) node[left=20pt] {} node[left]
     {\footnotesize {\text}} -- +(1.0, 0);

% ^1F levels      
  \node at (6.5,7.69) {\scriptsize $2s4f$};
  \node at (6.5,8.48) {\scriptsize $2s5f$};
  \node at (6.5,8.80) {\scriptsize $2s6f$};
  \node at (6.5,9.20) {\scriptsize $2s7f$};
  \coordinate (F00) at (6, 7.89);
  \coordinate (F10) at (6, 8.68);
  \coordinate (F20) at (6, 9.00);
  \coordinate (F30) at (6, 9.40);

  % Draw main levels
  \foreach \level/\text in {00/4, 10/5, 20/6, 30/7}
    \draw[level] (F\level) node[left=20pt] {} node[left]
     {\footnotesize {\text}} -- +(1.0, 0);

  % ^1G levels          
  \node at (8.5,8.46) {\scriptsize $2s5g$};
  \node at (8.5,8.80) {\scriptsize $2s6g$};
  \node at (8.5,9.20) {\scriptsize $2s7g$};
  \coordinate (G00) at (8,8.66);
  \coordinate (G10) at (8,9.00);
  \coordinate (G20) at (8,9.40);

  % Draw main levels
  \foreach \level/\text in {00/5, 10/6, 20/7}
    \draw[level] (G\level) node[left=20pt] {} node[left]
     {\footnotesize {\text}} -- +(1.0, 0);

  % ^1H levels          
  \node at (10.5,8.80) {\scriptsize $2s6h$};
  \node at (10.5,9.20) {\scriptsize $2s7h$};
  \coordinate (H00) at (10,9.0);
  \coordinate (H10) at (10,9.4); 

  % Draw main levels
  \foreach \level/\text in {00/6, 10/7}
    \draw[level] (H\level) node[left=20pt] {} node[left]
     {\footnotesize {\text}} -- +(1.0, 0);

  % ^1I levels 
  \node at (12.5,9.20) {\scriptsize $2s7i$};
%  \node at (12.5,9.28) {\scriptsize $8i$};
  \coordinate (I00) at (12,9.40);

  % Draw main levels
  \foreach \level/\text in {00/7}
    \draw[level] (I\level) node[left=20pt] {} node[left]
    {\footnotesize {\text}} -- +(1.0, 0);

  % K levels

  \node at (14.5,9.40){\scriptsize $2s8k$};
  \coordinate (K00) at (14,9.60);

  % Draw main levels
  \foreach \level/\text in {00/8}
    \draw[level] (K\level) node[left=20pt] {} node[left]
      {\footnotesize {\text}} -- +(1.0, 0);

  % Ionization level
  \draw[ionization] (0, 10.0) node[left=20pt] {E(${\rm F}^{6+}$)}-- +( 15.0, 0);

  % Rydberg levels
  \draw[level] (0,9.6) node[left=20pt] {}-- +(1.0, 0);
  \draw[level] (0,9.7) node[left=20pt] {}-- +(1.0, 0);
  \draw[level] (0,9.8) node[left=20pt] {}-- +(1.0, 0);
  \draw[level] (0,9.9) node[left=20pt] {}-- +(1.0, 0); 

  \draw[level] (2,9.6) node[left=20pt] {}-- +(1.0, 0);
  \draw[level] (2,9.7) node[left=20pt] {}-- +(1.0, 0);
  \draw[level] (2,9.8) node[left=20pt] {}-- +(1.0, 0);
  \draw[level] (2,9.9) node[left=20pt] {}-- +(1.0, 0);

  \draw[level] (4,9.6) node[left=20pt] {}-- +(1.0, 0);
  \draw[level] (4,9.7) node[left=20pt] {}-- +(1.0, 0);
  \draw[level] (4,9.8) node[left=20pt] {}-- +(1.0, 0);
  \draw[level] (4,9.9) node[left=20pt] {}-- +(1.0, 0);

  \draw[level] (6,9.6) node[left=20pt] {}-- +(1.0, 0);
  \draw[level] (6,9.7) node[left=20pt] {}-- +(1.0, 0);
  \draw[level] (6,9.8) node[left=20pt] {}-- +(1.0, 0);
  \draw[level] (6,9.9) node[left=20pt] {}-- +(1.0, 0);

  \draw[level] (8,9.6) node[left=20pt] {}-- +(1.0, 0);
  \draw[level] (8,9.7) node[left=20pt] {}-- +(1.0, 0);
  \draw[level] (8,9.8) node[left=20pt] {}-- +(1.0, 0);
  \draw[level] (8,9.9) node[left=20pt] {}-- +(1.0, 0);

  \draw[level] (10,9.6) node[left=20pt] {}-- +(1.0, 0);
  \draw[level] (10,9.7) node[left=20pt] {}-- +(1.0, 0);
  \draw[level] (10,9.8) node[left=20pt] {}-- +(1.0, 0);
  \draw[level] (10,9.9) node[left=20pt] {}-- +(1.0, 0);

  \draw[level] (12,9.7) node[left=20pt] {}-- +(1.0, 0);
  \draw[level] (12,9.8) node[left=20pt] {}-- +(1.0, 0);
  \draw[level] (12,9.9) node[left=20pt] {}-- +(1.0, 0);

  \draw[level] (14,9.8) node[left=20pt] {}-- +(1.0, 0);
  \draw[level] (14,9.9) node[left=20pt] {}-- +(1.0, 0);

  \end{tikzpicture}
\end{center}

\vspace{1cm}

\caption{The energy levels of the triplet states in the F$^{5+}$ ion. The threshold
energy (or ionization limit) $E($F$^{6+})$ =-82.330 336 543 $a.u.$ coincides with the total energy of the
ground 2$^2S$-state of the three-electron F$^{6+}$ ion.}

\end{sidewaysfigure}
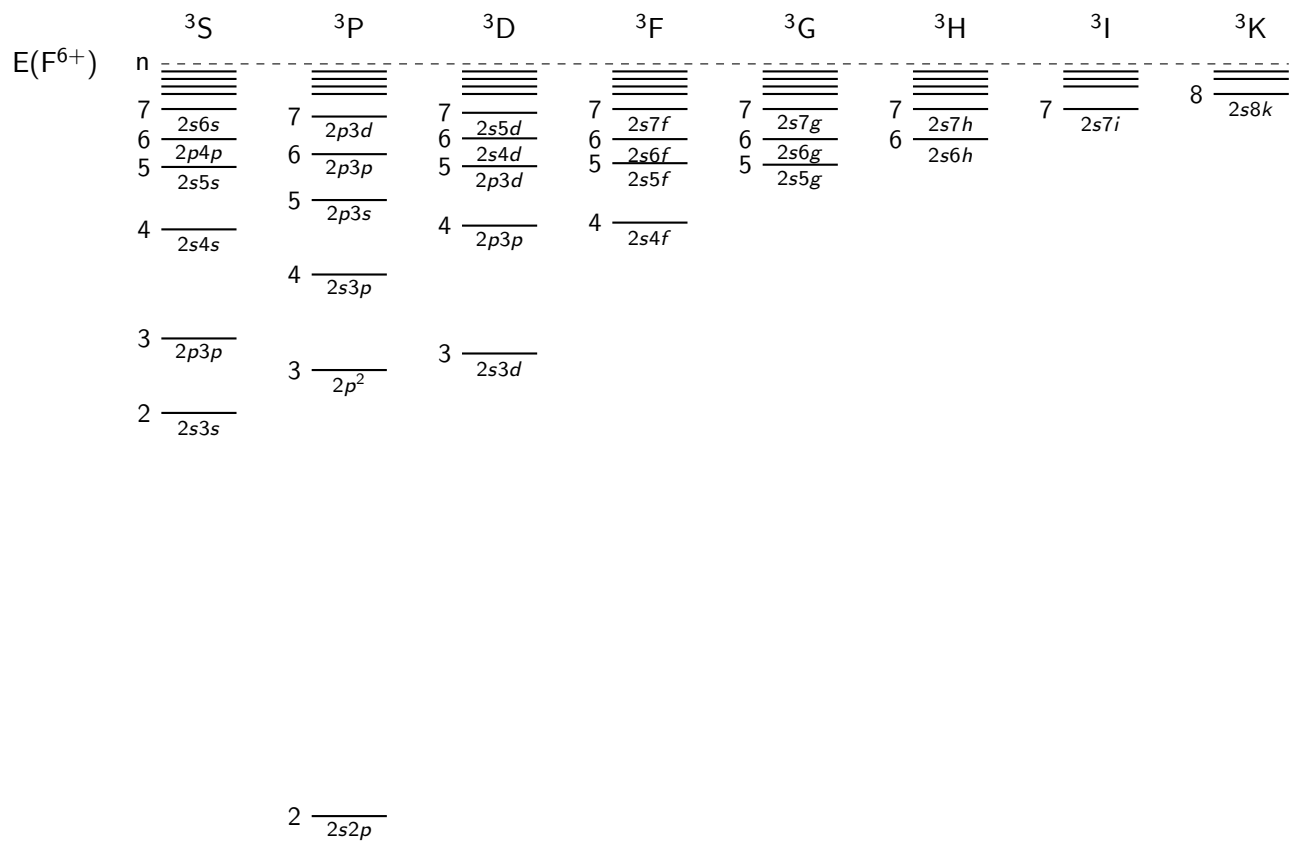

\newpage

% PIC1 Mg8+ ION triplet states

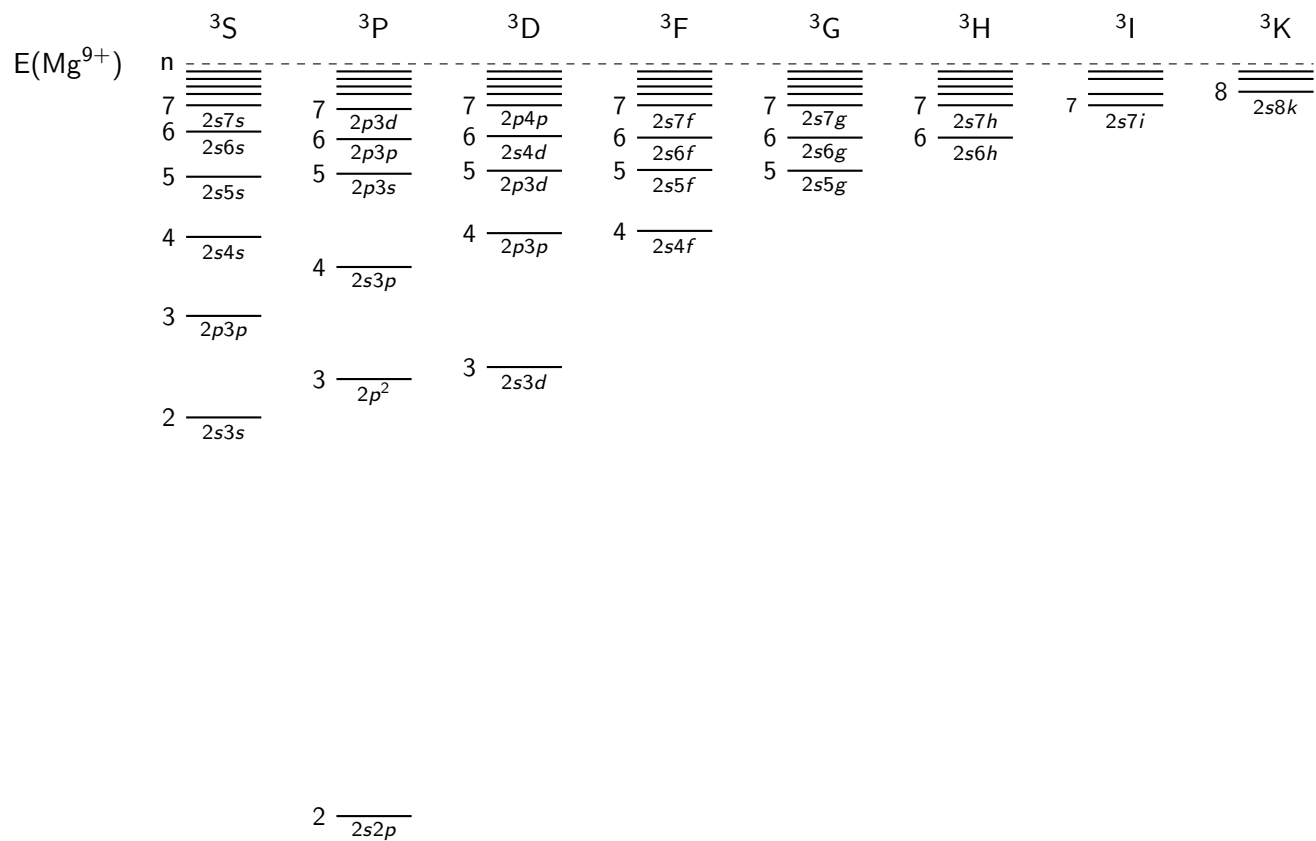
\begin{sidewaysfigure}

\begin{center}
  \sansmath
  \begin{tikzpicture}[
    font=\sffamily,
    level/.style={black,thick},
    ionization/.style={black,dashed},scale=1.0
  ]
  \coordinate (sublevel) at (0, 8pt);

  \node at (-0.25,10.0) {n};
  \node at (0.5, 10.5){$^3$S};
  \node at (2.5, 10.5){$^3$P}; 
  \node at (4.5, 10.5){$^3$D};
  \node at (6.5, 10.5){$^3$F}; 
  \node at (8.5, 10.5){$^3$G};
  \node at (10.5, 10.5){$^3$H};
  \node at (12.5, 10.5){$^3$I}; 
  \node at (14.5, 10.5){$^3$K};

  % S levels
  \node at (0.5, 5.10) {\scriptsize $2s3s$};
  \node at (0.5, 6.45) {\scriptsize $2p3p$};
  \node at (0.5, 7.50) {\scriptsize $2s4s$};
  \node at (0.5, 8.30) {\scriptsize $2s5s$};
  \node at (0.5, 8.90) {\scriptsize $2s6s$};
  \node at (0.5, 9.25) {\scriptsize $2s7s$};
% \node at (0.5, 8.30) {\scriptsize $2s8s$};

  \coordinate (S00) at (0, 5.30);
  \coordinate (S10) at (0, 6.65);
  \coordinate (S20) at (0, 7.70);
  \coordinate (S30) at (0, 8.50);
  \coordinate (S40) at (0, 9.10);
  \coordinate (S50) at (0, 9.45);
% \coordinate (S60) at (0, 8.50);

  % Draw main levels
%  \foreach \level/\text in { 00/2,  10/3,  20/4,  30/5}
   \foreach \level/\text in { 00/2,  10/3,  20/4,  30/5, 40/6, 50/7}
     \draw[level] (S\level) node[left=20pt] {} node[left]
      {\footnotesize {\text}} -- +(1.0, 0);

  % P levels          

  % ^1P levels          
  \node at (2.5,-0.20) {\scriptsize $2s2p$};
  \node at (2.5,5.61) {\scriptsize $2p^2$};
  \node at (2.5,7.10) {\scriptsize $2s3p$};
  \node at (2.5,8.34) {\scriptsize $2p3s$};
  \node at (2.5,8.80) {\scriptsize $2p3p$};
  \node at (2.5,9.20) {\scriptsize $2p3d$};
%  \node at (2.5,9.42) {\scriptsize $8p$};

  \coordinate (P00) at (2, 0.0 );
  \coordinate (P10) at (2, 5.81);
  \coordinate (P20) at (2, 7.30);
  \coordinate (P30) at (2, 8.54);
  \coordinate (P40) at (2, 9.00);
  \coordinate (P50) at (2, 9.40);

  % Draw main levels
   \foreach \level/\text in {00/2, 10/3, 20/4, 30/5, 40/6, 50/7}
%  \foreach \level/\text in {00/2, 10/3, 20/4}
    \draw[level] (P\level) node[left=20pt] {} node[left]
     {\footnotesize {\text}} -- +(1.0, 0);

% % ^1D levels          
  \node at (4.5,5.77) {\scriptsize $2s3d$};
  \node at (4.5,7.55) {\scriptsize $2p3p$};
  \node at (4.5,8.38) {\scriptsize $2p3d$};
  \node at (4.5,8.84) {\scriptsize $2s4d$};
  \node at (4.5,9.25) {\scriptsize $2p4p$};
  \coordinate (D00) at (4, 5.97);
  \coordinate (D10) at (4, 7.75);
  \coordinate (D20) at (4, 8.58);
  \coordinate (D30) at (4, 9.04);
  \coordinate (D40) at (4, 9.45);

% % Draw main levels
   \foreach \level/\text in {00/3, 10/4, 20/5, 30/6, 40/7}           
% \foreach \level/\text in {00/3, 10/4}
    \draw[level] (D\level) node[left=20pt] {} node[left]
     {\footnotesize {\text}} -- +(1.0, 0);

% ^1F levels      
  \node at (6.5,7.58) {\scriptsize $2s4f$};
  \node at (6.5,8.39) {\scriptsize $2s5f$};
  \node at (6.5,8.82) {\scriptsize $2s6f$};
  \node at (6.5,9.25) {\scriptsize $2s7f$};
  \coordinate (F00) at (6, 7.78);
  \coordinate (F10) at (6, 8.59);
  \coordinate (F20) at (6, 9.02);
  \coordinate (F30) at (6, 9.45);

  % Draw main levels
  \foreach \level/\text in {00/4, 10/5, 20/6, 30/7}
    \draw[level] (F\level) node[left=20pt] {} node[left]
     {\footnotesize {\text}} -- +(1.0, 0);

  % ^1G levels          
  \node at (8.5,8.38) {\scriptsize $2s5g$};
  \node at (8.5,8.82) {\scriptsize $2s6g$};
  \node at (8.5,9.25) {\scriptsize $2s7g$};
  \coordinate (G00) at (8,8.58);
  \coordinate (G10) at (8,9.02);
  \coordinate (G20) at (8,9.45); 

  % Draw main levels
  \foreach \level/\text in {00/5, 10/6, 20/7}
    \draw[level] (G\level) node[left=20pt] {} node[left]
     {\footnotesize {\text}} -- +(1.0, 0);

  % ^1H levels          
  \node at (10.5,8.82) {\scriptsize $2s6h$};
  \node at (10.5,9.25) {\scriptsize $2s7h$};
  \coordinate (H00) at (10,9.02);
  \coordinate (H10) at (10,9.45);

  % Draw main levels
  \foreach \level/\text in {00/6, 10/7}
    \draw[level] (H\level) node[left=20pt] {} node[left]
     {\footnotesize {\text}} -- +(1.0, 0);

  % ^1I levels 
  \node at (12.5,9.25) {\scriptsize $2s7i$};
%  \node at (12.5,9.28) {\scriptsize $8i$};
  \coordinate (I00) at (12,9.45);

  % Draw main levels
  \foreach \level/\text in {00/7}
    \draw[level] (I\level) node[left=20pt] {} node[left]
    {\scriptsize {\text}} -- +(1.0, 0);

  % K levels
 
  \node at (14.5,9.43){\scriptsize $2s8k$};
  \coordinate (K00) at (14,9.63);

  % Draw main levels
  \foreach \level/\text in {00/8}
    \draw[level] (K\level) node[left=20pt] {} node[left]
      {\footnotesize {\text}} -- +(1.0, 0);

  % Ionization level
  \draw[ionization] (0, 10.0) node[left=20pt] {E(${\rm Mg}^{9+}$)}-- +( 15.0, 0);

  % Rydberg levels
  \draw[level] (0,9.6) node[left=20pt] {}-- +(1.0, 0);
  \draw[level] (0,9.7) node[left=20pt] {}-- +(1.0, 0);
  \draw[level] (0,9.8) node[left=20pt] {}-- +(1.0, 0);
  \draw[level] (0,9.9) node[left=20pt] {}-- +(1.0, 0); 

  \draw[level] (2,9.6) node[left=20pt] {}-- +(1.0, 0);
  \draw[level] (2,9.7) node[left=20pt] {}-- +(1.0, 0);
  \draw[level] (2,9.8) node[left=20pt] {}-- +(1.0, 0);
  \draw[level] (2,9.9) node[left=20pt] {}-- +(1.0, 0);

  \draw[level] (4,9.6) node[left=20pt] {}-- +(1.0, 0);
  \draw[level] (4,9.7) node[left=20pt] {}-- +(1.0, 0);
  \draw[level] (4,9.8) node[left=20pt] {}-- +(1.0, 0);
  \draw[level] (4,9.9) node[left=20pt] {}-- +(1.0, 0);

  \draw[level] (6,9.6) node[left=20pt] {}-- +(1.0, 0);
  \draw[level] (6,9.7) node[left=20pt] {}-- +(1.0, 0);
  \draw[level] (6,9.8) node[left=20pt] {}-- +(1.0, 0);
  \draw[level] (6,9.9) node[left=20pt] {}-- +(1.0, 0);

  \draw[level] (8,9.6) node[left=20pt] {}-- +(1.0, 0);
  \draw[level] (8,9.7) node[left=20pt] {}-- +(1.0, 0);
  \draw[level] (8,9.8) node[left=20pt] {}-- +(1.0, 0);
  \draw[level] (8,9.9) node[left=20pt] {}-- +(1.0, 0);

  \draw[level] (10,9.6) node[left=20pt] {}-- +(1.0, 0);
  \draw[level] (10,9.7) node[left=20pt] {}-- +(1.0, 0);
  \draw[level] (10,9.8) node[left=20pt] {}-- +(1.0, 0);
  \draw[level] (10,9.9) node[left=20pt] {}-- +(1.0, 0);

  \draw[level] (12,9.6) node[left=20pt] {}-- +(1.0, 0);
  \draw[level] (12,9.8) node[left=20pt] {}-- +(1.0, 0);
  \draw[level] (12,9.9) node[left=20pt] {}-- +(1.0, 0);

  \draw[level] (14,9.8) node[left=20pt] {}-- +(1.0, 0);
  \draw[level] (14,9.9) node[left=20pt] {}-- +(1.0, 0);

  \end{tikzpicture}
\end{center}

\vspace{1cm}

\caption{The energy levels of the triplet states in the Mg$^{8+}$ ion. The threshold
energy (or ionization limit) $E($Mg$^{9+})$ =-150.136 154 391 $a.u.$ coincides with the total energy of the
ground 2$^2S$-state of the three-electron Mg$^{9+}$ ion.}

\end{sidewaysfigure}

\newpage

% TABLE I

\begin{table}[tp]
\begin{center}
\caption{List of of the different $L$ configurations used in the CI calculations of the $S$, $P$, $D$, $F$, $G$, $H$, $I$, and $K$
states. In all configurations $M=0$. The exponents within a shell have been kept equal.}
\begin{tabular}{|c | c | c | c |}
\hline\hline
State & $L$ & $M$ & Configurations \\
\hline
$S$  & 0 & 0 & $ssss$, $sspp$, $spps$, $ppss$, $pppp$, $ssdd$, $sdds$, $ddss$, $sppd$, $dpps$, $sdpp$, \\
     &   &   & $ppds$, $ppdd$, $dddd$, $ddpp$, $ssff$, $ppff$, $ddff$, $ffss$, $ffpp$, $ffdd$ \\
\hline
$P$  & 1 & 0 & $sssp$, $spss$, $sppp$, $ppsp$, $sddp$, $ddsp$, $sffp$, $ffsp$, $sspd$, $spds$, $pdss$, \\
     &   &   & $pdpp$, $pddd$, $pdff$, $pppd$, $ddpd$, $ffpd$, $ssdf$, $ppdf$, $dddf$, $ffdf$, $pdpp$ \\
\hline
$D$  & 2 & 0 & $sssd$, $sdss$, $sppd$, $ppsd$, $sddp$, $ddsd$, $sffd$, $ffsd$,  \\
     &   &   & $sspp$, $spps$, $ppss$, $pppp$, $ssdd$, $sdds$, $ddss$  \\
\hline
$F$  & 3 & 0 & $sssf$, $sfss$, $sppf$, $ppsf$, $sddf$, $ddsf$, $ffsf$, $ggsf$, $hhsf$ \\
\hline
$G$  & 4 & 0 & $sssg$, $sgss$, $sppg$, $ppsg$, $sddg$, $ddsg$, $ffsg$, $ggsg$, $hhsg$ \\
\hline
$H$  & 5 & 0 & $sssh$, $shss$, $spph$, $ppsh$, $sddh$, $ddsh$, $ffsh$, $ggsh$, $hhsh$  \\
\hline
$I$  & 6 & 0 & $sssi$, $siss$, $sppi$, $ppsi$, $sddi$, $ddsi$, $ffsi$, $ggsi$, $hhsi$, $iisi$ \\
\hline
$K$  & 7 & 0 & $sssk$, $skss$, $sppk$, $ppsk$, $sddk$, $ddsk$, $ffsk$, $ggsk$, $hhsk$, $kksk$ \\
\hline\hline
\end{tabular}
\end{center}
\end{table}

% TABLE II

\begin{table}[tp]
\begin{center}
\caption{The total energies in $a.u.$ of some low-lying singlet states of the ${}^{\infty}$Be-atom determined
         with the use of the CI method and comparison with the non-relativistic values of the bibliography.
         The total energies of the bound states in the upper part of this Table
         are lower than the ionization threshold of the ${}^{\infty}$Be atom ($E_{{}^{\infty}{\rm Be}^{+}}$ =
         -14.324 763 176 790 43(22) $a.u.$ \cite{PKP}). N is the number of configurations used in calculations.
         Diff. are the energy differences between the present and reference energies in milli-Hartree ($1\cdot 10^{-3}$ a.u.).}
    \scalebox{0.85}{%
\begin{tabular}{| c  | l | l  |  l | l | c | l | l | c | c |}
\hline\hline
 State &  N   &   This work      & N  &   $E(CI,MC)$    &     Ref.     & N  &     $E(nr)$     & Ref.  & Diff. \\
\hline
2$^1$S & 1137 & -14.665 730 & $\approx 2m$& -14.667 347 30 & \cite{Bunge}  &  4096 & -14.667 356 4949  & \cite{PKP13}  & 1.63 \\
3$^1$S & 1300 & -14.416 247 &        1038 & -14.417 957 27 & \cite{Chung}  & 10000 & -14.418 240 328   & \cite{AdamS}  & 2.00 \\
4$^1$S & 1466 & -14.365 107 &             &                &               & 10000 & -14.370 087 876   & \cite{AdamS}  & 4.98 \\
5$^1$S & 1242 & -14.358 945 &             &                &               & 10000 & -14.351 511 654   & \cite{AdamS}  & 2.45 \\
6$^1$S & 1358 & -14.340 351 &             &                &               & 10000 & -14.342 403 552   & \cite{AdamS}  & 2.05 \\
\hline
2$^1$P & 1307 & -14.470 359 &        1038 & -14.473 009 65 & \cite{Chung}  & 10700 & -14.473 451 378   & \cite{AdamP}  & 3.09 \\
3$^1$P & 1398 & -14.389 310 &        1038 & -14.392 788 28 & \cite{Chung}  & 11600 & -14.393 143 528   & \cite{AdamP}  & 3.83 \\
4$^1$P & 1944 & -14.353 697 &             &                &               & 11900 & -14.361 938 388   & \cite{AdamP}  & 8.24 \\
5$^1$P & 2159 & -14.326 239 &             &                &               & 12200 & -14.347 876 275   & \cite{AdamP}  & 21.64 \\
\hline
3$^1$D &  943 & -14.401 671 &             & -14.404 36(6)  & \cite{Galv}   &  4200 & -14.408 237 03(40)& \cite{AdamD}  & 6.57 \\
4$^1$D & 1754 & -14.367 016 &        1038 & -14.373 442 41 & \cite{Chung}  &  4200 & -14.373 824 38(30)& \cite{AdamD}  & 6.81 \\
5$^1$D & 1131 & -14.343 314 &      &              &                        &  4200 & -14.353 982 65(50)& \cite{AdamD}  &10.67 \\
6$^1$D &      & -14.319 243 &      &              &                        &  4200 & -14.343 857 75(60)& \cite{AdamD}  &24.61 \\
\hline
4$^1$F & 2210 & -14.352 803 &      &              &               &       &                 &                     &      \\
5$^1$F & 1625 & -14.334 737 &      &              &               &       &                 &                     &      \\
\hline
5$^1$G & 1676 & -14.338 416 &      &              &               &       &                 &                     &      \\
6$^1$G & 1844 & -14.318 712 &      &              &               &       &                 &                     &      \\
\hline
6$^1$H & 1443 & -14.327 148 &      &              &               &       &                 &                     &      \\
\hline
7$^1$I & 1242 & -14.319 606 &      &              &               &       &                 &                     &      \\
\hline
8$^1$K & 1540 & -14.285 509 &      &              &               &       &                 &                     &     \\
\hline\hline
\end{tabular}}
\end{center}
\end{table}

\newpage

\begin{table}[tp]
\begin{center}
\caption{CI calculations on the triplet excited states of Be-atom and isoelectronic ions. Energy in a.u.}
\scalebox{0.85}{
\begin{tabular}{| c  | c | c | c | c | c |}
\hline\hline
 State &     E(Be)    &  E(B$^+$)   &  E(C$^{2+}$) &    E(F$^{5+}$)     &   E(Mg$^{8+}$)    \\
\hline
2$^3S$ & -14.428 858  & -23.756 406  & -35.448 470 & -84.697 926 & -155.200 640  \\
3$^3S$ & -14.371 277  & -23.589 166  & -35.123 226 & -84.002 052 & -154.198 338  \\
4$^3S$ & -14.343 789  & -23.517 737  & -35.061 579 & -83.959 916 & -154.131 317  \\
5$^3S$ & -14.314 690  & -23.473 813  & -35.037 812 & -83.592 771 & -152.883 937   \\
6$^3S$ &              & -23.457 241  & -34.978 838 & -83.115 745 & -152.029 948  \\
7$^3S$ &              & -23.437 292  & -34.885 511 & -82.994 777 & -152.014 389   \\
8$^3S$ &              & -23.329 741  & -34.775 418 & -82.975 972 & -151.847 275  \\
\hline
2$^3P$ & -14.565 432  & -24.176 888  & -36.294 390 & -87.659 827 & -161.532 045  \\
3$^3P$ & -14.392 598  & -23.687 970  & -35.347 035 & -84.498 979 & -154.905 474  \\
4$^3P$ & -14.357 020  & -23.559 695  & -35.127 064 & -84.132 562 & -154.393 494  \\
5$^3P$ & -14.333 088  & -23.515 121  & -35.081 574 & -83.881 773 & -154.017 858  \\
6$^3P$ &              & -23.502 078  & -34.989 957 & -83.709 305 & -153.901 853  \\
7$^3P$ &              & -23.428 174  & -34.948 115 & -83.798 028 & -153.881 813  \\
8$^3P$ &              &              & -34.936 695 & -83.494 189 & -152.728 404  \\
\hline
3$^3D$ & -14.381 020  & -23.651 575  & -35.290 221 & -84.383 794 & -154.733 038     \\
4$^3D$ & -14.350 318  & -23.548 131  & -35.060 046 & -83.474 253 & -152.702 810     \\
5$^3D$ & -14.343 322  & -23.493 956  & -34.948 953 & -83.052 943 & -151.756 880     \\
6$^3D$ &              & -23.445 075  & -34.869 323 & -82.802 203 & -151.228 105     \\
7$^3D$ &              & -23.357 994  & -34.754 413 & -82.135 779 & -150.526 015     \\
8$^3D$ &              & -23.141 951  & -34.334 354 & -79.533 960 & -147.349 779     \\
\hline
4$^3F$ & -14.351 575  & -23.546 249  & -35.053 823 & -83.454 221 & -152.661 603     \\
5$^3F$ & -14.334 657  & -23.490 973  & -34.941 051 & -83.036 317 & -151.737 664     \\
\hline
5$^3G$ & -14.336 099  & -23.489 411  & -34.941 295 & -83.043 091 & -151.749 511    \\
6$^3G$ & -14.335 897  & -23.387 245  & -34.810 143 & -82.746 257 & -151.139 034    \\
\hline
6$^3H$ & -14.326 506  & -23.464 042  & -34.888 913 & -82.817 790 & -151.248 300     \\
\hline
7$^3I$ & -14.320 140  & -23.440 907  & -34.853 261 & -82.670 858 & -150.914 473     \\
\hline
8$^3K$ & -14.315 243  & -23.420 425  & -34.821 733 & -82.560 657 & -150.729 034    \\
\hline\hline
\end{tabular}}
\end{center}
\footnotetext[1]{The ionization threshold of the ${}^{\infty}$Be atom ($E_{{}^{\infty}{\rm Be}^{+}}$ =-14.324 763 176 790 43(22) $a.u.$ \cite{PKP}).}
\footnotetext[2]{The ionization threshold of the ${}^{\infty}$Be atom ($E_{{}^{\infty}{\rm B}^{2+}}$ =-23.424 605 665     $a.u.$).}
\footnotetext[3]{The ionization threshold of the ${}^{\infty}$Be atom ($E_{{}^{\infty}{\rm C}^{3+}}$ =-34.775 510 611    $a.u.$).}
\footnotetext[4]{The ionization threshold of the ${}^{\infty}$Be atom ($E_{{}^{\infty}{\rm F}^{6+}}$ =-82.330 336 543 $a.u.$).}
\footnotetext[5]{The ionization threshold of the ${}^{\infty}$Be atom ($E_{{}^{\infty}{\rm Mg}^{9+}}$ =-150.136 154 391 $a.u.$).}
\end{table}

\newpage

% TABLE III

\begin{table}[tp]
\begin{center}
\caption{Order of bound states in the Be-atom from the theoretical calculations and experimental observations of spectral lines \cite{Kramida}. The symbol '$\textless$' denotes here lower energy.}
%    \scalebox{0.85}{%
\begin{tabular}{| c |  c | c |}
\hline\hline
$n$ &   order in singlet series              &  order in triplet series \\
\hline
   &    $^2$Be$^{+}$  Ionization Limit      &  $^2$Be$^{+}$  Ionization Limit      \\
\hline
13 & 13$^1$D $\textless$ 13$^1$P                                            & 13$^3$D \\
12 & 12$^1$D $\textless$ 12$^1$P                                            & 12$^3$D \\
11 & 11$^1$D $\textless$ 11$^1$S $\textless$ 11$^1$P                        & 11$^3$D \\
10 & 10$^1$D $\textless$ 10$^1$S $\textless$ 10$^1$P                        & 10$^3$D  \\
9  &  9$^1$D $\textless$ 9$^1$S $\textless$ 9$^1$P                          & 9$^3$D   \\
8  &  8$^1$D $\textless$ 8$^1$S $\textless$ 8$^1$P                          & 8$^3$P    \\
7  &  7$^1$D $\textless$ 7$^1$S $\textless$ 7$^1$P $\textless$ 7$^1$F $\textless$ 7$^1$G $\textless$ 7$^1$H $\textless$ 7$^1$I   & 7$^3$D $\textless$ 7$^3$F $\textless$ 7$^3$S \\
6  &  6$^1$D $\textless$ 6$^1$S $\textless$ 6$^1$P $\textless$ 6$^1$F $\textless$ 6$^1$G $\textless$ 6$^1$H  & 6$^3$P  $\textless$ 6$^3$D $\textless$ 6$^3$F $\textless$ 6$^3$S \\
5  &  5$^1$D $\textless$ 5$^1$S $\textless$ 5$^1$P $\textless$ 5$^1$F $\textless$ 5$^1$G   & 5$^3$P  $\textless$ 5$^3$D $\textless$ 5$^3$F $\textless$ 5$^3$G $\textless$ 5$^3$S \\
4  &  4$^1$D $\textless$ 4$^1$S $\textless$ 4$^1$P $\textless$ 4$^1$F      & 4$^3$P  $\textless$ 4$^3$D $\textless$ 4$^3$F $\textless$ 4$^3$S \\
3  &  3$^1$S $\textless$ 3$^1$D $\textless$ 3$^1$P             & 3$^3$P,3$^3P^{\prime}$$^a$ $\textless$ 3$^3$D $\textless$ 3$^3$S   \\
2  &  2$^1$S $\textless$ 2$^1$P                                & 2$^3$P $\textless$ 2$^3$S        \\
\hline
\end{tabular}
%}
\end{center}
\footnotetext[1]{Here we denote the doublet of states with same $n$ energy level.}
\end{table}

\end{document}